\documentclass[aps,prd,showpacs,twocolumn]{revtex4}

\RequirePackage[T1]{fontenc}

\usepackage{graphicx}
\usepackage{amssymb}
\usepackage{epstopdf}
\usepackage{hyperref}
\usepackage{amsmath}
\usepackage{bbold}
\usepackage{color}
\usepackage{ulem}
\usepackage[title]{appendix}

\begin{document}

\title{Exploring Direct Detection Suppressed Regions in a Simple 2-Scalar Mediator Model of Scalar Dark Matter}

\author{J\'er\^ome Claude\footnote{Email: jerome.claude@carleton.ca} and Stephen Godfrey\footnote{Email: godfrey@physics.carleton.ca}}
%\thankstext{e1}{e-mail: jerome.claude@carleton.ca}
%\thankstext{e2}{e-mail: godfrey@physics.carleton.ca}
\affiliation{
Ottawa-Carleton Institute for Physics, 
Department of Physics, Carleton University, Ottawa, Canada K1S 5B6 \label{addr1}
}

\date{\today}

\begin{abstract}
We explore regions of parameter space that give rise to suppressed 
direct detection cross sections in 
a simple model of scalar dark matter with a scalar portal
that mixes with the standard model Higgs. We found that even  this simple model allows  
considerable room in the parameter space that has not been excluded by direct detection limits.
A number of effects leading to this result have been previously noted. Our main new result 
explores interference effects between different contributions to DM annihilation when the DM
mass is larger than the scalar portal mass. New annihilation channels open up 
and the parameters of the model need to compensate to
give the correct DM relic abundance, resulting in smaller direct detection cross sections.
We find that even in a very simple model
of DM there are still sizeable regions of parameter space that are not ruled out by experiment.
\end{abstract}

\maketitle

\section{Introduction}

There is considerable evidence for {\it Dark Matter} (DM), a type of matter in the universe which has so far 
only revealed itself through gravitational interactions with normal matter~\cite{Bergstrom:2000pn,Bertone:2004pz,Bergstrom:2012fi}.
DM at most interacts
very weakly with normal matter. Various means of DM interacting with normal matter have been explored;
Higgs portals, 
e.g.~\cite{Bertone:2004pz,Silveira:1985rk,McDonald:1993ex,Patt:2006fw,Baek:2011aa,Djouadi:2011aa,LopezHonorez:2012kv,Baek:2012se,Walker:2013hka,Esch:2014jpa,Buchmueller:2014yoa,Cheung:2015dta,Beniwal:2015sdl,Han:2016gyy,Arcadi:2016qoz,Arcadi:2017kky,Bhattacharya:2017fid,Gross:2017dan,Azevedo:2018oxv,Arcadi:2019lka,Cabrera:2019gaq,Ghorbani:2014gka,Ghosh:2017fmr,Alanne:2020jwx}, 
vector portals, e.g.~\cite{Beniwal:2015sdl,Arcadi:2017kky,Buckley:2011mm,Frandsen:2011cg,Lebedev:2011iq,Alves:2013tqa,Arcadi:2013qia,Lebedev:2014bba,Arcadi:2014lta,Hooper:2014fda,Alves:2015pea,Ghorbani:2015baa,Jacques:2016dqz,Duerr:2016tmh,Ismail:2016tod,Escudero:2016gzx,Kearney:2016rng,Alves:2016cqf,Dutra:2018gmv,Okada:2018ktp,Blanco:2019hah,Arcadi:2020jqf,Okada:2020cue}, 
and neutrino portals, e.g.~\cite{Beniwal:2015sdl,Arcadi:2017kky,Falkowski:2009yz,Cherry:2014xra,Batell:2017cmf,Cosme:2020mck}. 
Higgs portal models have been tightly constrained by experiment,
leaving only small 
regions in the parameter space viable~\cite{Escudero:2016gzx}. In particular, 
direct detection experiments have tightly constrained
the parameter space. 
However, there still exist allowed regions, including regions
referred to as {\it blind spots} which are 
due to cancellations in the direct detection cross section amplitudes.
This has been explored in a number of papers, 
for example~\cite{Baek:2012se,Gross:2017dan,Azevedo:2018oxv,Cabrera:2019gaq,Cheung:2012qy,Cheung:2013dua,Huang:2014xua,Berlin:2015wwa,Casas:2017jjg,Huitu:2018gbc}.
In addition to the blind spots mechanism, other mechanisms exist that 
 suppress direct detection cross sections which we discuss below.

Many Higgs portal models have a second
scalar that mixes with the Standard Model (SM) Higgs 
boson~\cite{Baek:2011aa,LopezHonorez:2012kv,Baek:2012se,Esch:2014jpa,Cheung:2015dta,Arcadi:2016qoz,Gross:2017dan,Azevedo:2018oxv,Arcadi:2019lka,Cabrera:2019gaq,Ghosh:2017fmr,Alanne:2020jwx,Duerr:2016tmh,Huang:2014xua,Barger:2007im,Falkowski:2015iwa,Ko:2016ybp,Arcadi:2016kmk,Bell:2016ekl,Bell:2017rgi,Arcadi:2018pfo}. 
The mechanism leading to blind spots in such models
is the destructive interference between the Higgs-like scalar and the second scalar in the direct detection
cross section 
amplitude~\cite{LopezHonorez:2012kv,Baek:2012se,Esch:2014jpa,Arcadi:2016qoz,Gross:2017dan,Azevedo:2018oxv,Cabrera:2019gaq,Alanne:2020jwx,Huang:2014xua,Arcadi:2016kmk,Bell:2017rgi}. 
Given that detecting dark matter is the focus of a broad
experimental program, we felt it useful to further 
explore regions of the parameter space that give rise to suppressed 
direct detection cross sections. Our preconceived bias
was that the mixing angle between the two $t$-channel exchange bosons could be tuned to create the
direct detection blind spots mentioned above. However, we found that values of the mixing angle
that would give rise to blind spots are for the most part ruled out by measurements of Higgs boson properties --- 
most generally by the Higgs signal strengths, but also by the Higgs invisible width
when Higgs decay to dark matter is kinematically allowed.
Another mechanism that can lead to suppressed direct detection cross sections
which has previously been pointed out~\cite{Esch:2014jpa,Arcadi:2017kky}
is the result of a resonance effect occurring when the dark matter mass is roughly half
the value of the scalar portal mass. However, 
there is a third mechanism that 
suppresses the direct detection cross section when 
the dark matter particle is more massive than either the Higgs boson or the portal 
particle\footnote{Ref.~\cite{Esch:2014jpa,Azevedo:2018oxv,Ghosh:2017fmr} 
noted a similar effect when $m_S > m_{h_1,h_2}$ although the 
details differ from those presented here.}. In this 
case, a large region of the parameter space has not been ruled out by any of the theory constraints, 
any experimental constraints and, more to the point of this exercise, by direct
detection limits.

For the purposes of this study, 
we constructed a very simple toy model consisting of scalar DM and an additional scalar portal that can mix
with the SM Higgs field to study direct detection suppressed regions. 
We use this toy model to explore effects  for the simplest possible case 
of a scalar dark matter portal extension.
There are, however, many possible variations of this simple picture that can
give rise to cancellations in the direct detection cross section. An 
incomplete list of possibilities appearing in the literature consists of the scalar portal being replaced with a 
pseudoscalar portal~\cite{Arcadi:2019lka,Berlin:2015wwa,Buckley:2014fba,No:2015xqa,Goncalves:2016iyg,Tunney:2017yfp,Arcadi:2017wqi,Ghosh:2020fdc,Butterworth:2020vnb,Abe:2020pzp}, 
or having a complex scalar which gives rise to a  second scalar portal~\cite{Gross:2017dan,Alanne:2020jwx,Huitu:2018gbc,Okada:2020zxo,Alanne:2020jwx,Zhang:2021alu}, 
 a two Higgs doublet model~\cite{Arcadi:2019lka,Cabrera:2019gaq,Berlin:2015wwa,Bell:2017rgi,Arcadi:2018pfo,Arcadi:2017wqi,Ghosh:2020fdc,Butterworth:2020vnb,Zhang:2021alu,Bauer:2017ota,Jiang:2019soj,Arcadi:2020gge}, 
higher Higgs representations~\cite{Arhrib:2011uy},
 or supersymmetric models~\cite{Arcadi:2019lka,Cheung:2012qy,Huang:2014xua,Ellis:2005mb,Carena:2006nv,Hooper:2006wv,Cao:2010ph}.
Before proceeding, we note that, given that we simply want to push the simplest of models
as far as we could, we haven't dealt with the issue of UV completeness.  
However, Gross, Lebedev and Toma~\cite{Gross:2017dan} and Huitu {\it et. al.}~\cite{Huitu:2018gbc}
showed that they could make models very similar to ours UV complete by assuming the system is invariant under a 
global $U(1)$ which is gauged in the UV-completeness. Other examples of similar UV complete models
are \cite{Ghosh:2017fmr,Alanne:2020jwx}.

 Our simple  model has eight parameters but two are fixed to their SM values, one is fixed to give the correct
DM relic abundance, and one is only weakly constrained by DM self-interactions. This leaves four
independent parameters which we choose to be the scalar DM mass, the scalar portal mass, the 
scalar singlet vacuum expectation value, and the mixing angle between the SM Higgs scalar and the scalar
portal. We scan through the parameter space and, by transforming to the Lagrangian parameters,
we test that perturbative unitarity holds, that the potential is bounded from below, and that the parameters
result in a consistent set of parameters for the desired properties of the model.
We next fix the remaining parameter to
give the correct relic abundance. With these parameter values, we test that the parameters are consistent with 
the Higgs boson invisible width, the Higgs signal strengths, and DM self-interaction limits. 
Finally, we calculate the direct and indirect detection cross sections and compare them to the
experimental limits. 

In Section~\ref{sec:model}, we give the details of our model and examine the theoretical constraints on its 
parameters. 
In Section~\ref{sec:scan}, we describe the details of scanning the parameter space and the various
experimental measurements we use to constrain parameter points, starting by fitting
the DM-portal coupling to the DM relic abundance. The remaining experimental 
constraints are the Higgs invisible width, the Higgs 
signal strength, the DM self-interaction, and the DM indirect detection cross section. 
We then compare the points
that pass all these constraints to the direct detection limits and examine the various
mechanisms that lead to direct detection suppressed regions.
Finally, in Section~\ref{sec:conclusions}, we summarize our conclusions.

\section{A 2-Scalar Mediator Model with Scalar DM}
\label{sec:model}
We consider an extension of the Standard Model that consists of two singlet scalar fields 
$\varphi$ and $S$, with $\varphi$ a portal particle that mixes with the SM Higgs field
and $S$ the DM particle.
We impose a $Z_2 \times Z_2 $ symmetry on these fields so that they
are odd under their respective $Z_2$'s to ensure their 
stability and eliminate terms
in the potential odd in $\varphi$ and $S$ (see for example Ref.~\cite{Arcadi:2017kky,Arcadi:2019lka}). 
We note that the $Z_2$ imposed on $\varphi$
is spontaneously broken when $\varphi$ acquires a vev.
The most general scalar potential with this symmetry is then given by

\begin{eqnarray}
V(H, \varphi, S) & = & -\mu_H^2 H^\dagger H + \lambda_H (H^\dagger H)^2 \nonumber\\
& - & {{\mu_\varphi^2}\over 2}\varphi^2 + {{\lambda_\varphi}\over 4}\varphi^4 + \lambda_4 \varphi^2 (H^\dagger H )\nonumber \\
& + & {{\mu_S^2}\over 2} S^2 + {{\lambda_S}\over 4} S^4 + {{ \lambda_{\varphi\varphi SS}}\over 2} \varphi^2 S^2 \nonumber\\
& + & {{\lambda_{HH SS}}\over 2} (H^\dagger H) S^2.
\label{eqn:V}
\end{eqnarray}
Following Ref.~\cite{Arcadi:2017kky,Arcadi:2019lka}, we take $\lambda_{HH SS} = 0$ 
so that the Standard Model complex scalar doublet $H$ does not 
directly couple to the dark matter candidate, $S$, at tree level. 
This choice does not affect our conclusions, and we will discuss the consequences
of not taking $\lambda_{HH SS} = 0$ in Section~\ref{sec:lnez} after we present our results.
This term can be generated via $\varphi$ loops and the natural size for the resulting vertex would 
be the product of the couplings $ \lambda_{\varphi\varphi SS} \lambda_4  /(16\pi^2)$. We assume 
that the vertex can be made small enough even if it requires some amount of tuning.
Assuming this term is small enough, and because the DM thermally averaged annihilation cross section
is typically dominated by the $s$-channel annihilation cross section and real production of $h_2$, we will find
that it will not have a big effect on the relic abundance and that neglecting it 
will not qualitatively alter our conclusions.

We work in the unitarity gauge and shift the fields to the new minimum; 
$H \to (0, (v+h)/\sqrt{2})^T$ and $\varphi \to (w+\phi)$, where $v$ and $w$ 
are the vacuum expectation values (vevs) of the neutral component of $H$ and $\phi$ respectively. 
We require that $S$ does not acquire a vev so that the $Z_2$ symmetry remains unbroken and $S$
is stable.
With this substitution, we then minimize the 
resulting potential $V(h, \phi, S)$ with respect to the scalar fields and obtain
$\mu_H^2=\lambda_H v^2+\lambda_4w^2$ and $\mu_\varphi^2=\lambda_\varphi w^2+\lambda_4v^2$. After substituting
these expressions into $V(h, \phi, S)$, we find the mass terms from the resulting potential.
Diagonalizing the mass matrix for the 
$h$ and $\phi$ fields leads to physical states that are linear combinations of the
the $h$ and $\phi$ fields with mixing angle $\alpha$ given by:
\begin{eqnarray}
h_1 	& = 	& h\cos{\alpha}-\phi\sin{\alpha} \\
h_2 	& = 	& \phi\cos{\alpha}+h\sin{\alpha}
\end{eqnarray}
with 
\begin{eqnarray}
\sin(2\alpha)	& =	& \frac{2\lambda_4vw}{\sqrt{(\lambda_H v^2 - \lambda_\varphi w^2)^2 +4\lambda_4^2 v^2w^2}} \\
\cos(2\alpha)	& = & \frac{\lambda_\varphi w^2-\lambda_H v^2}{\sqrt{(\lambda_H v^2 - \lambda_\varphi w^2)^2 +4\lambda_4^2 v^2w^2}},
\end{eqnarray}
and the scalar masses given by
\begin{eqnarray}
m_{h_1}^2 	& = 	& \lambda_H v^2+\lambda_\varphi w^2-\frac{\lambda_\varphi w^2-\lambda_H v^2}{\cos{(2\alpha)}} \\
m_{h_2}^2	& = 	& \lambda_H v^2+\lambda_\varphi w^2+\frac{\lambda_\varphi w^2-\lambda_H v^2}{\cos{(2\alpha)}} \\
m_S^2		& =	& \mu_S^2+\lambda_{\varphi\varphi SS} w^2.
\end{eqnarray}
For small values of $\alpha$, we identify $h_1$ with the $125~\text{GeV}$ scalar 
associated with the Standard Model Higgs boson. Because of the mixing, both $h_1$ and $h_2$ act as 
portals between the Standard Model and the dark matter candidate $S$.

When we scan the parameter space, we will use the physical parameters 
$m_{h_1}$, $m_{h_2}$, $\alpha$, $v$, and $w$,
but the theoretical constraints described below constrain the Lagrangian parameters. 
We will therefore need the relationships between the physical and the Lagrangian parameters, which are given by
\begin{eqnarray}
\lambda_H & = & \frac{1}{4v^2}\left(\left(m_{h_1}^2+m_{h_2}^2\right)-\left(m_{h_2}^2-m_{h_1}^2\right)\cos{2\alpha}\right) \label{eqn:lambdaH} \\
\lambda_\varphi & = & \frac{1}{4w^2}\left(\left(m_{h_1}^2+m_{h_2}^2\right)+\left(m_{h_2}^2-m_{h_1}^2\right)\cos{2\alpha}\right) \label{eqn:lambdaphi}\\
\lambda_4 & = & \frac{\sin{2\alpha}}{4vw}\left(m_{h_2}^2-m_{h_1}^2\right). \label{eqn:lambda4}
\end{eqnarray}

In the following subsections, we examine the theoretical constraints on the Lagrangian parameters.

\subsection{Constraints from Partial Wave Unitarity}

We start by using partial wave unitarity (PWU) of the $2\rightarrow2$ scattering amplitudes to 
constrain the Lagrangian parameters.
In the high energy limit, only tree level diagrams involving four-point scalar interactions
contribute, as diagrams involving propagators are suppressed by the square of the collision
energy. Under these conditions, only the zeroth partial wave amplitude $a_0$ contributes to the $2\rightarrow2$ amplitudes
$\mathcal{M}$, so that the constraint $\left|a_0\right|< \frac{1}{2}$ corresponds to 
$\mathcal{M} < 8\pi$.
In the high energy limit, we can also use
the Goldstone equivalence theorem to replace the gauge bosons with the Goldstone bosons.

There are therefore six fields to consider in the scattering amplitudes: $S$, $\varphi$,
and the four Goldstone bosons $\eta^0$, $\eta^{0*}$, $\eta^+$, and $\eta^-$.
The PWU condition must be applied to each of the eigenvalues 
of the coupled-channel scattering matrix $\mathcal{M}$ for
all pairs of incoming and outgoing scalar fields.
Because the scalar potential is invariant under $SU(2)\times U(1)$ symmetry,
the scattering processes conserve electric charge and hypercharge, and can be classified
by the total electric charge ($Q$) and hypercharge ($Y$) of the incoming and outgoing states. 
$S$ and $\varphi$ are SM gauge singlets and the Goldstone bosons come from the
$SU(2)_L$ doublet with $Y=1$ (where $Q_{em}=T_3 +Y/2$).
A symmetry factor of $1/\sqrt{2}$ is included for each pair of identical
particles in the initial and final states. 

Starting with the $Q=2$ and $Y=2$ quantum numbers,
there is only one scattering channel, $\eta^+\eta^+\rightarrow\eta^+\eta^+$, 
which leads to the constraint 
\begin{equation}
\left|\lambda_H\right| < 4\pi. 
\label{eqn:lh}
\end{equation}
Likewise, the only scattering amplitude for $Q=1$ and $Y=0$ is 
$\eta^+{\eta^0}^* \rightarrow\eta^+{\eta^0}^*$, which yields the same constraint.

For $Q=0$ and $Y=1$, there is only the $\eta^0\varphi\rightarrow\eta^0\varphi$ 
scattering amplitude, leading to the constraint 
\begin{equation}
\left|\lambda_4\right| < 4\pi.
\label{eqn:l4}
\end{equation}
Likewise, the only scattering amplitude for $Q=1$ and $Y=1$ is 
$\eta^+\varphi\rightarrow\eta^+\varphi$, which yields the same constraint.

For the $Q=0$ and $Y=0$ quantum numbers, there are five states:
$\eta^0\eta^{0*}$, $\eta^+ \eta^-$, $\varphi\varphi$, $\varphi S$, and $SS$. This results in 
a $5 \times 5 $ scattering matrix consisting of a $4 \times 4 $ block and the 
$ \varphi S \rightarrow \varphi S$ channel. The $ \varphi S \rightarrow \varphi S$ channel
leads to the constraint 
\begin{equation}
\left| \lambda_{\varphi\varphi SS} \right|< 4\pi .
\label{eqn:lppss}
\end{equation} 
We can partially diagonalize 
the $4 \times 4 $ matrix into a $3 \times 3 $ matrix and a diagonal term. The diagonal
term leads to the constraint $\left|\lambda_H\right| < 4\pi$. To find the remaining
constraints, we diagonalize the $3 \times 3 $ matrix by taking its determinant 
and imposing that the roots of the resulting polynomial satisfy $\left|\text{Roots}\left(p(x)\right)\right| < 8\pi$, where
\begin{equation}
\begin{split}
p(x)=\left(x-3\lambda_S\right)\left(-4\lambda_4^2+\left(x-6\lambda_H\right)\left(x-3\lambda_\varphi\right)\right) \\
 -\left(x-6\lambda_H\right)\lambda_{\varphi\varphi SS}^2.
\end{split}
\end{equation}
We follow the procedure of Ref.~\cite{Campbell:2016zbp} to which we direct the interested reader
for details, and replace the bounds on the roots of $p(x)$
with the three equivalent conditions:
\begin{eqnarray}
16\pi & > & \left| 6\lambda_H + 3\lambda_\varphi \pm \sqrt{\left(6\lambda_H-3\lambda_\varphi\right)^2+16\lambda_4^2}\right| \label{eqn:z1} \\
\lambda_S & < & \frac{1}{3}\left[8\pi+\frac{\left(6\lambda_H-8\pi\right)\lambda_{\varphi\varphi SS}^2}{\left(6\lambda_H-8\pi\right)\left(3\lambda_\varphi-8\pi\right)-4\lambda_4^2}\right] \label{eqn:z2}\\
\lambda_S & > & \frac{1}{3}\left[-8\pi+\frac{\left(6\lambda_H+8\pi\right)\lambda_{\varphi\varphi SS}^2}{\left(6\lambda_H+8\pi\right)\left(3\lambda_\varphi+8\pi\right)-4\lambda_4^2}\right] \label{eqn:z3}.
\end{eqnarray}
Thus, the constraints on the Lagrangian parameters from partial wave unitarity are given
by equations \ref{eqn:lh},\ref{eqn:l4}, \ref{eqn:lppss}, \ref{eqn:z1}, \ref{eqn:z2} and \ref{eqn:z3}.

\subsection{Constraints from the Bounded from Below Requirement}

We next include constraints on the Lagrangian parameters that ensure
 that the scalar potential is bounded from below.
Because the quartic terms dominate at large field values, this constraint acts on
the quartic terms in the potential.

We use the approach described in Ref.~\cite{Arhrib:2011uy} (see also Ref.~\cite{Campbell:2016zbp})
in which we use a hyperspherical coordinate system replacing the scalar fields by the following parameters:
\begin{eqnarray}
r & =& \sqrt{ \left| H \right| ^2 + \varphi^2 + S^2} \\
r \sin{\beta}\cos{\gamma} & =& \left| H \right| ^2 \\
r \sin{\beta}\sin{\gamma} & =& \varphi^2 \\
r \cos{\beta} & =& S^2. \\
\end{eqnarray}
The quartic part of the potential can be then be written as
\begin{equation}
\frac{r^4}{(1+\tan^2{\beta})(1+\tan^2{\gamma})} \mathbf{x}^\intercal \mathbf{A} \mathbf{y}
\end{equation}
where
\begin{eqnarray}
A & = & \frac{1}{4}\begin{bmatrix}
\lambda_S & 2\lambda_S& \lambda_S \\
0 & 2\lambda_{\varphi\varphi SS} & 2\lambda_{\varphi\varphi SS} \\
4\lambda_H &4 \lambda_4 & \lambda_\varphi \\
\end{bmatrix}\\
x & = & \begin{bmatrix} 1 \\ \tan{\beta} \\ \tan^2{\beta} \end{bmatrix} \\
y & = & \begin{bmatrix} 1 \\ \tan{\gamma} \\ \tan^2{\gamma} \end{bmatrix}.
\end{eqnarray}
Since the prefactor is strictly positive, the requirement for the potential to be bounded from below 
is that $\mathbf{x}^\intercal \mathbf{A} \mathbf{y}$ be positive. 
This term can be written as a quadratic in $\tan^2{\beta}$ with factors themselves quadratics in $\tan^2{\gamma}$, 
or vice-versa. Requiring these expressions to be positive leads to the following constraints:
\begin{eqnarray}
\lambda_H & > & 0 \\
\lambda_\varphi & > & 0 \\
\lambda_S & > & 0 \\
\lambda_4 & > & -\sqrt{\lambda_H \lambda_\varphi} \\
\lambda_{\varphi\varphi SS} & > & -\sqrt{\lambda_\varphi \lambda_S}. \label{eqn:bblppss}
\end{eqnarray}

\subsection{Constraints from Consistency of the Potential}

With the sign conventions in our potential, for the $H$ and $\varphi$ fields to obtain a vev and for 
$S$ to not obtain a vev we require $\mu_H^2 >0$, $\mu_\varphi^2 >0$, and $\mu_S^2 >0$. This leads to the following 
three constraints:
\begin{eqnarray}
\mu_H^2 & = & \lambda_H v^2+\lambda_4 w^2 > 0 \label{eqn:muH}\\
\mu_\varphi^2 & = & \lambda_\varphi w^2+\lambda_4 v^2 > 0 \label{eqn:muphi}\\
\mu^2_S & = & m_S^2 - \lambda_{\varphi\varphi SS} w^2 >0. \label{eqn:muS}
\end{eqnarray}
Imposing these constraints gives the only consistent set of parameters with a DM candidate.
Under these conditions, the potential and minima are unique.

\section{Parameter Scan and Relic Abundance}
\label{sec:scan}
The model has eight independent parameters. At the Lagrangian level, these parameters 
are $\lambda_H$, $\lambda_\varphi$, $\lambda_4$, $\lambda_S$, $\lambda_{\varphi\varphi SS}$, 
$\mu_H$, $\mu_\varphi$, and $\mu_S$. 
However, it is more transparent to use more physical parameters. We take these to
be $m_{h_1}$, $m_{h_2}$, $m_S$, the $h$-$\varphi$ mixing angle $\alpha$, the two 
vacuum expectation values $v$ and $w$, and retain the Lagrangian parameters 
 $\lambda_{\varphi\varphi SS}$ and $\lambda_S$. The relationship between these and the
 Lagrangian parameters was given by Eqns.~\ref{eqn:lambdaH}, \ref{eqn:lambdaphi}, \ref{eqn:lambda4}, 
 \ref{eqn:muH}, \ref{eqn:muphi} and \ref{eqn:muS}.

We identify $v$ with the SM Higgs vacuum expectation value 
and $m_{h_1}$ with the observed 125~GeV scalar mass, leaving
six parameters. Of these, $\lambda_S$ is constrained by 
dark matter self-interaction and Eqns.~\ref{eqn:z2} and \ref{eqn:z3}. 
 When these two constraints are not mutually exclusive,
 $\lambda_S$ can be set to an arbitrary value that satisfies these constraints
 without impacting any other quantity of interest. 
 $\lambda_{\varphi\varphi SS}$ directly influences the dark matter annihilation cross section, and
 we fix its value to give agreement with the measured relic abundance after all other 
 parameters have been fixed. This leaves $m_{h_2}$, $m_S$, $\alpha$, and $w$ as free input parameters.
 
Our procedure is to randomly choose values for $m_{h_2}$, $m_S$, $\alpha$, and $w$. 
We can limit
the allowed range on $\alpha$ using the measured Higgs boson signal strengths to constrain
$|\cos\alpha| \gtrsim 0.97$. This will be checked later by comparing the calculated and measured
signal strengths. We typically scan the four parameters by randomly 
varying $w$ and $m_S$ from $1~\text{GeV}$ to $1~\text{TeV}$, 
$m_{h_2}$ from $100~\text{GeV}$ to $1~\text{TeV}$,
and $\alpha$ from 
$0.969 < |\cos\alpha| < 1.0 $. We take $\lambda_S=0.2$. 
We note that we find no qualitative difference in our results or conclusions by
increasing the scan range for $m_S$, $m_{h_2}$, and $w$ to larger values
so that scanning to 1~TeV is sufficient to reveal 
the characteristics we are exploring.

We then check the resulting Lagrangian
parameters against the relevant theoretical constraints. For the parameter sets that pass
this test, we 
use the micrOMEGAs program~\cite{Belanger:2018ccd} to search for values of 
$\lambda_{\varphi\varphi SS}$ that agree with the measured value for the relic 
abundance of $\Omega_\text{DM}=0.1200(12) \; h^{-2}$~\cite{Tanabashi:2018oca}. We then check the Lagrangian parameters
against the remaining theoretical constraints. For those that pass this test, we calculate
and compare to experimental measurements
the Higgs boson invisible branching ratio, the Higgs boson signal strength, and the 
dark matter self-interaction cross section. For those 
parameter points that pass all these constraints, we calculate the indirect detection 
cross sections for all possible final states and the direct detection cross section
using micrOMEGAs~\cite{Belanger:2018ccd}. The 
goal is to see if parameter points that pass all these theoretical and experimental tests
are either ruled out or allowed by current limits on direct and indirect detection 
cross section measurements. 

In the following subsections, we describe the details of how we do these calculations.

\subsection{Fitting $\lambda_{\varphi\varphi SS}$ with the Relic Abundance}
\label{sec:fitting}

\begin{figure}
\centering
\includegraphics[width=0.48\textwidth]{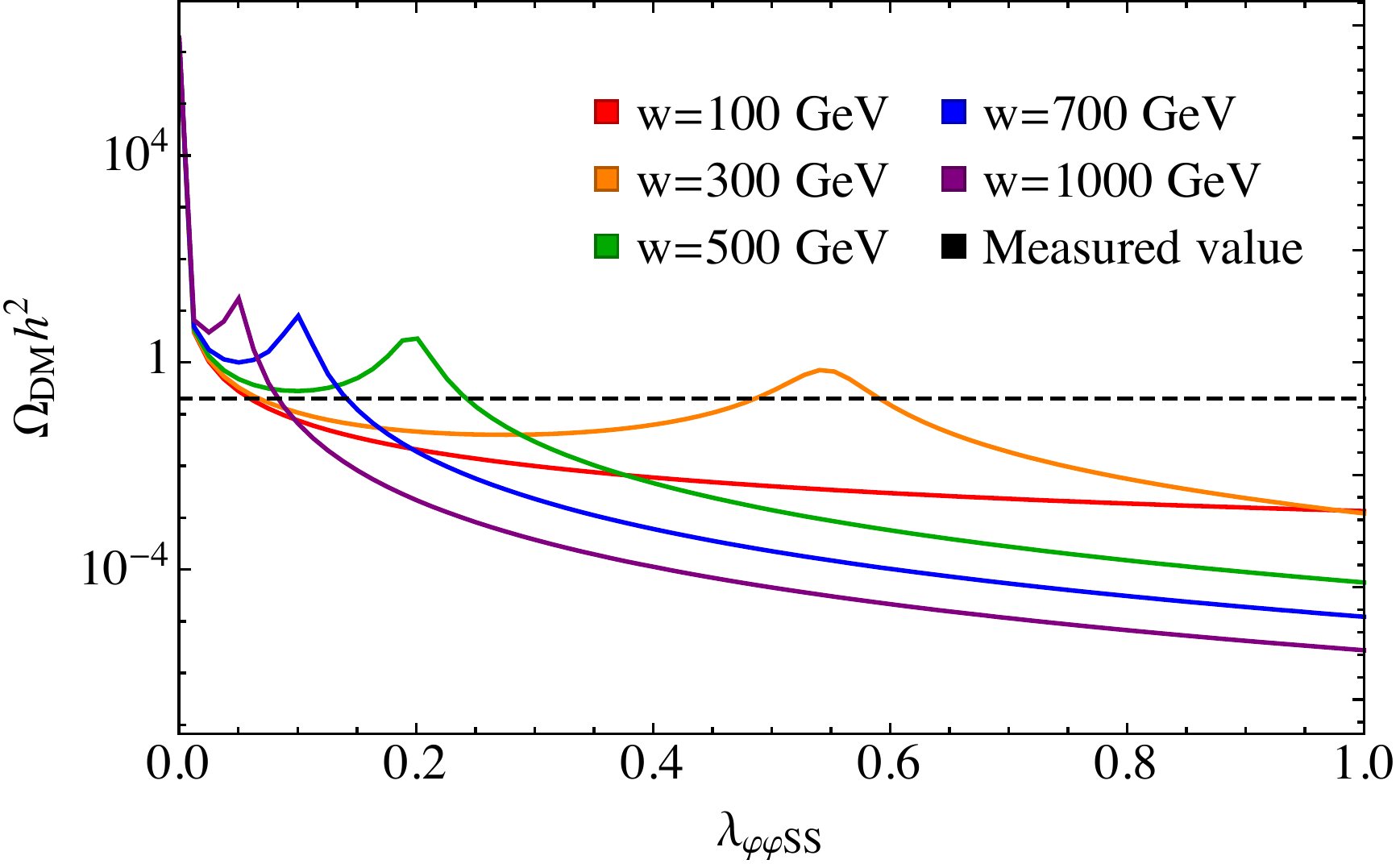}
\caption{Dark matter relic abundance as a function of $\lambda_{\varphi\varphi SS}$ for $\alpha=0.2$, $m_{h_2}=200~\text{GeV}$, $m_S=300~\text{GeV}$, and for 
the fixed values of $w$ given in the legend. The dashed line is for the measured value
of $\Omega_\text{DM} =0.1200(12)\; h^{-2}$~\cite{Tanabashi:2018oca}. }
\label{fig:lamOmega}
\end{figure}

We calculate the relic abundance and other DM properties using the micrOMEGAs program~\cite{Belanger:2018ccd}.
For each set of input parameters, we perform a search by 
varying $\lambda_{\varphi\varphi SS}$ until we obtain agreement between
the calculated value for $\Omega_\text{DM}$ and the measured value. However, when $m_S \gtrsim m_{h2}$,
the relic abundance is no longer a monotonic function of $\lambda_{\varphi\varphi SS}$,
which complicates the search and can lead to up to three solutions. For these cases,
the relic abundance starts by decreasing with increasing $\lambda_{\varphi\varphi SS}$
but then increases again due to a cancellation in the DM annihilation cross sections. This
is illustrated in Fig.~\ref{fig:lamOmega}. 

\begin{figure}
\centering
\includegraphics[width=0.48\textwidth]{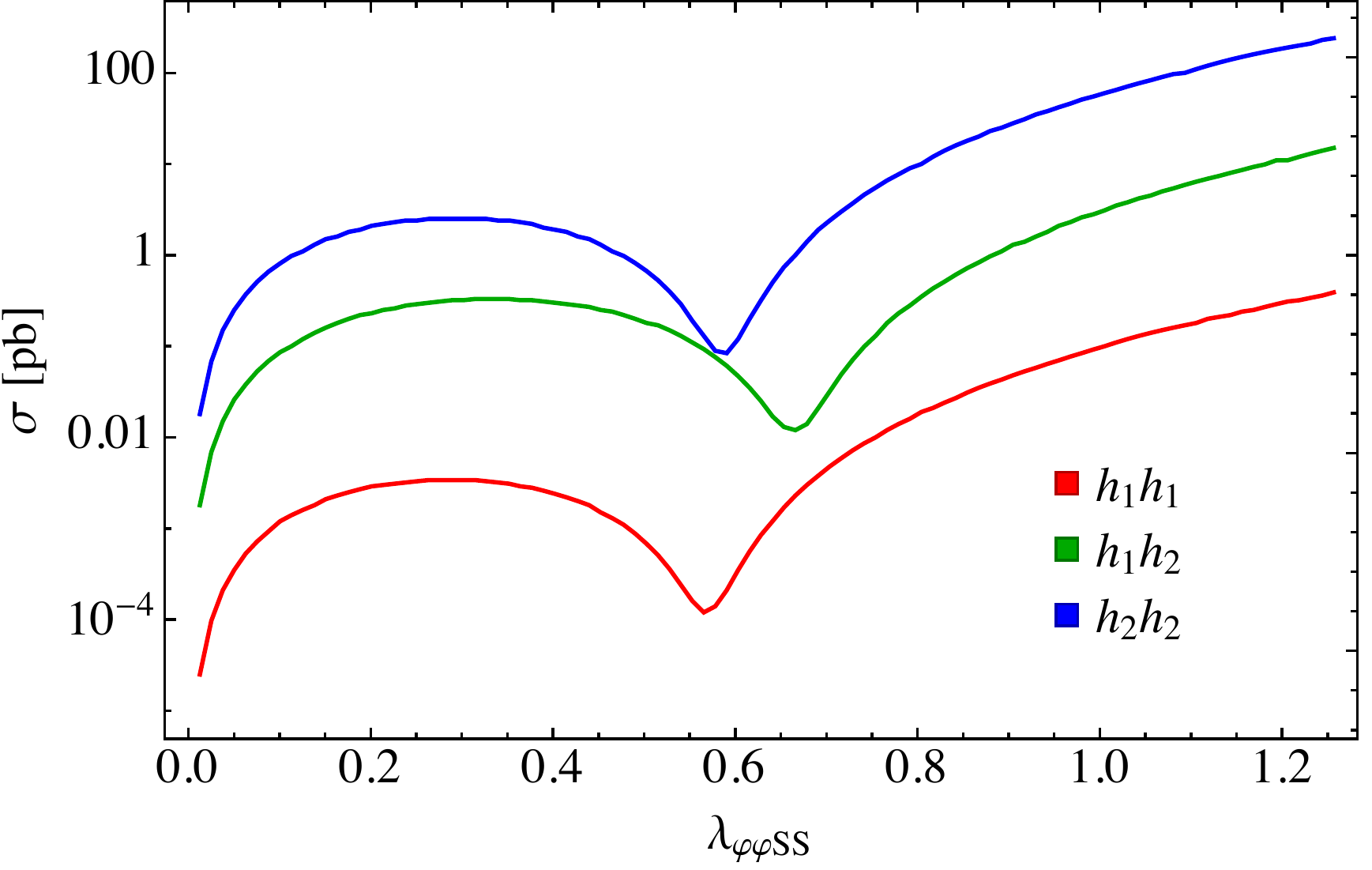}
\caption{The dark matter annihilation cross section to scalar channels as a function of 
$\lambda_{\varphi\varphi SS}$ for $\alpha=0.2$, $w=300~\text{GeV}$, $m_{h_2}=200~\text{GeV}$, 
and $m_S=300~\text{GeV}$, for a center of mass energy of $125~\text{GeV}$. 
The value of $\lambda_{\varphi\varphi SS}$ where the cross section 
is a minimum is different for each channel.}
\label{fig:sigmapoles}
\end{figure}

The cancellation is due to interference between the
diagrams  contributing to the $SS \rightarrow h_2 h_2$ cross section that,
for small $h_1$-$h_2$ mixing angles, 
occurs at $\lambda_{\varphi\varphi SS} \approx m_S^2/2w^2$. This is 
a consequence of the Feynman rules for the various vertices entering these matrix elements; 
 the details are presented in Appendix~\ref{appendix}.
As is well known, when the annihilation cross section decreases, the relic abundance increases due 
to earlier freeze-out. 
For finite values of the mixing angle,
this effect is also present in the $h_1h_2$ and $h_1h_1$ final states, 
although it occurs at different values of $\lambda_{\varphi\varphi SS}$ for each channel; 
this can be seen in Fig.~\ref{fig:sigmapoles}. 
While the $h_2h_2$ final state generally dominates because
the $h_1h_2$ and $h_1h_1$ are suppressed by factors of
$\sin\alpha$ and $\sin^2\alpha$ respectively, 
all channels contribute to the relic abundance so that there is no simple formula for
the location of the maximum in $ \Omega_\text{DM}$. As a consequence, we use
the small mixing angle formula given above to approximate the position of the maxima.
While the value of $m_S$ only affects the amplitude of the maxima, 
$\alpha$ does influence their position, so this formula is not very accurate for large values 
of $\alpha$. Nonetheless, the formula is an adequate 
approximation for the local maximum in $ \Omega_\text{DM}$ for the purposes of searching for
the values of $\lambda_{\varphi\varphi SS}$ that give the correct relic abundance
value $\Omega_\text{DM}=0.1200(12)\; h^{-2}$~\cite{Tanabashi:2018oca}.

In general, as pointed out above, there can be up to three values of 
$\lambda_{\varphi\varphi SS}$ that give $\Omega_\text{DM}=0.12\; h^{-2}$. We must therefore 
take some care in our search so that we do not miss one of these solutions.
For $m_S<200$~GeV, the maximum is not high enough to yield additional solutions 
for $\lambda_{\varphi\varphi SS}$. It is therefore sufficient to 
perform a simple search procedure starting at 
$\lambda_{\varphi\varphi SS}=0$. 
From this starting point, we increase $\lambda_{\varphi\varphi SS}$ in small increments until 
$\Omega_\text{DM}$ falls below 
$0.12 \; h^{-2}$, after which we perform a binary search between the last two values 
of $\lambda_{\varphi\varphi SS}$ until we find a value of $\lambda_{\varphi\varphi SS}$
that yields $\Omega_\text{DM}=0.12\; h^{-2}$. If this does not occur before 
$\lambda_{\varphi\varphi SS}$ reaches $4\pi$, the scan is aborted.

For larger values of $m_S$, we determine the position of the maximum in $ \Omega_\text{DM}$ using 
$\lambda_{\varphi\varphi SS}^\text{max} = m_S^2/2w^2$. If $\Omega_\text{DM} < 0.12 \; h^{-2}$ for 
$\lambda_{\varphi\varphi SS}^\text{max}$, there are no additional solutions due to the maximum, 
and we follow the procedure described above starting at $\lambda_{\varphi\varphi SS}=0$ to 
determine the unique solution, if it exists.

If $\Omega_\text{DM} > 0.12\; h^{-2}$, we follow the procedure described 
above starting at $\lambda_{\varphi\varphi SS}^\text{max}$ to find a solution to the right of 
the maximum. We repeat this procedure, this time
decreasing $\lambda_{\varphi\varphi SS}$ from $\lambda_{\varphi\varphi SS}^\text{max}$
to find a solution to the left of the maximum. If one is found, the procedure is repeated 
starting from $\lambda_{\varphi\varphi SS}=0$ to find the final solution.

This yields a list of points in the parameter space that give the correct relic abundance.
We then check to see that the values of $\lambda_{\varphi\varphi SS}$
satisfy the remaining theoretical constraints given by Eqns.~\ref{eqn:z2},\ref{eqn:z3},
\ref{eqn:bblppss} and \ref{eqn:muS}. 

Once we have a set of parameters that give the correct relic abundance and satisfy the theoretical 
constraints, we test them against the experimental constraints.

\subsection{Constraints from the Higgs Invisible Width}

The current limits on the invisible width of the $H^0$ boson at 125~GeV is 
$\text{BR}_\text{invis}<0.26$ at 95\% C.L. (ATLAS~\cite{Aaboud:2019rtt}) and
$\text{BR}_\text{invis}<0.19$ at 95\% C.L. (CMS~\cite{Sirunyan:2018owy}).
We use the less constraining limit of $\text{BR}_\text{invis}<0.26$ but this 
has very little effect on our results.
Identifying $h_1$ with the $H^0$, the $h_1$ invisible BR is given by
\begin{equation}
\text{BR}_\text{inv}=\frac{\Gamma_\text{inv}}{\Gamma_\text{total}}=\frac{\Gamma_\text{inv}}{\Gamma_\text{SM}\cos^2\alpha +\Gamma_\text{inv}}
\label{eqn:BRinv}
\end{equation}
where $\Gamma_\text{SM} = 4.07$~GeV~\cite{Tanabashi:2018oca} (see also HDECAY~\cite{Djouadi:1997yw}),
which is modified by the $h_1$-$h_2$ mixing, $\cos\alpha$.
The $h_1 SS$ vertex is $2i\lambda_{\varphi\varphi SS} w \sin{\alpha}$, so that
the invisible width is given by 
\begin{equation}
\Gamma_\text{inv}=\frac{\lambda_{\varphi\varphi SS}^2 w^2 \sin^2{\alpha}}{8\pi m_{h_1}}\sqrt{1-4\frac{m_S^2}{m_{h_1}^2}}.
\end{equation}
This constraint eliminates points for $m_S \lesssim m_{h_1}/2$, where the kinematically 
allowed decay $h_1 \to SS$ results in a large $\Gamma_\text{inv}$.

\subsection{Constraints from the Higgs Signal Strength}

The Higgs signal strength $\mu$ is given by
\begin{equation}
\mu=\sum_i c_i \omega_i,
\end{equation}
where the sum runs over all channels, and where the channel signal strength $c_i$ 
and the SM channel weight $\omega_i$ are given by
\begin{eqnarray}
c_i & = & \frac{\left[\sigma \times \text{BR}\right]_i}{\left[\sigma_\text{SM} \times \text{BR}_\text{SM}\right]_i} \\
\omega_i & = & \frac{\epsilon_i\left[\sigma_\text{SM} \times \text{BR}_\text{SM}\right]_i}{\sum_j\epsilon_j\left[\sigma_\text{SM} \times \text{BR}_\text{SM}\right]_j}
\end{eqnarray}
for channel $i$ with cross section $\sigma$ ($\sigma_\text{SM}$) 
and branching ratio $\text{BR}$ ($\text{BR}_\text{SM}$) in the BSM (SM) 
model and $\epsilon_i$ the experimental efficiency for that channel~\cite{Bechtle:2013xfa}. 
For the Standard Model, the Higgs signal strength parameter is $\mu=1$. 
The current PDG quoted average for the signal strength is $\mu = 1.13 \pm .06$~\cite{Tanabashi:2018oca}.
In our model, $\mu \leq 1$. As such, relative to this best fit point, the 95\% C.L. 
limit is $\mu>0.94$.

For our model, all production channels are modified equally by the $h_1$-$h_2$ mixing, 
$\cos\alpha$. This leads to a factor of $\sigma_i /\sigma_{\text{SM}i}=\cos^2{\alpha}$ for
the production channels. The decay channels are slightly different as one needs to include the 
modification of the invisible width in the total width, so that 
\begin{equation}
\frac{\text{BR}_i}{\text{BR}_{\text{SM}i}}=\frac{\Gamma_{\text{SM}}}{\Gamma_{\text{SM}i}}\frac{\Gamma_i}{\Gamma}
= \cos^2{\alpha} \frac{\Gamma_{\text{SM}}}{\Gamma}
\end{equation}
where $\Gamma = \Gamma_\text{SM}\cos^2\alpha +\Gamma_\text{inv}$.
Putting it together we obtain
\begin{equation}
\mu=\cos^4{\alpha}\frac{\Gamma_\text{SM}}{\Gamma},
\end{equation}
which can be used to apply the constraint $\mu>0.94$.
This constraint eliminates parameter points for which $\cos\alpha\lesssim 0.97$, as anticipated. 
Additionally,
when $h_1$ is kinematically allowed to decay to $SS$, the $h_1$ width is significantly 
larger than the Standard Model value so that the signal strength is altered, also eliminating
parameter points.

\subsection{Constraints from Dark Matter Self-Interaction}

At tree level, the strength of dark matter self-interaction is determined by 
$\lambda_S$ from the quartic coupling and $\lambda_{\varphi\varphi SS}$ 
from $t$-channel and $s$-channel processes. Once $\lambda_{\varphi\varphi SS}$ 
is set by the relic abundance, we compare the predicted self-interaction cross section to 
current limits on $\sigma_\text{DM}$.
Constraints from the Bullet Cluster give a limit of $\sigma_\text{SIDM}/m_S < 1.25$~cm$^2$/g~\cite{Robertson:2016xjh}, while constraints from colliding galaxies clusters give
$\sigma_\text{DM}/m < 0.47$~cm$^2$/g (95\% CL)~\cite{Harvey:2015hha}.
We use the tighter constraint of $\sigma_\text{SIDM}/m_S < 0.47$~cm$^2$/g~$\approx
2.2 \times 10^3$~GeV$^{-3}$. However, $\sigma_\text{SIDM}$ 
only constrains $\lambda_S$, and we have chosen a value that avoids this limit.

\subsection{Constraints from Indirect Detection}
\label{sec:ID}
Dwarf spheroidal satellite galaxies (dSphs) are typically DM dominated, and so are a good 
place to study dark matter. We calculated cross-sections for our model 
using the micrOMEGAs program~\cite{Belanger:2018ccd} which outputs $\sigma_\text{ID} v$ at rest. We 
compared our results to 
a global analysis by Hoof {\it et al}~\cite{Hoof:2018hyn}
of DM signals from 27 dwarf spheroidal galaxies using 11 years of 
observations by Fermi-LAT~\cite{Atwood:2009ez}. 

\begin{figure}
\centering
\includegraphics[width=0.5\textwidth]{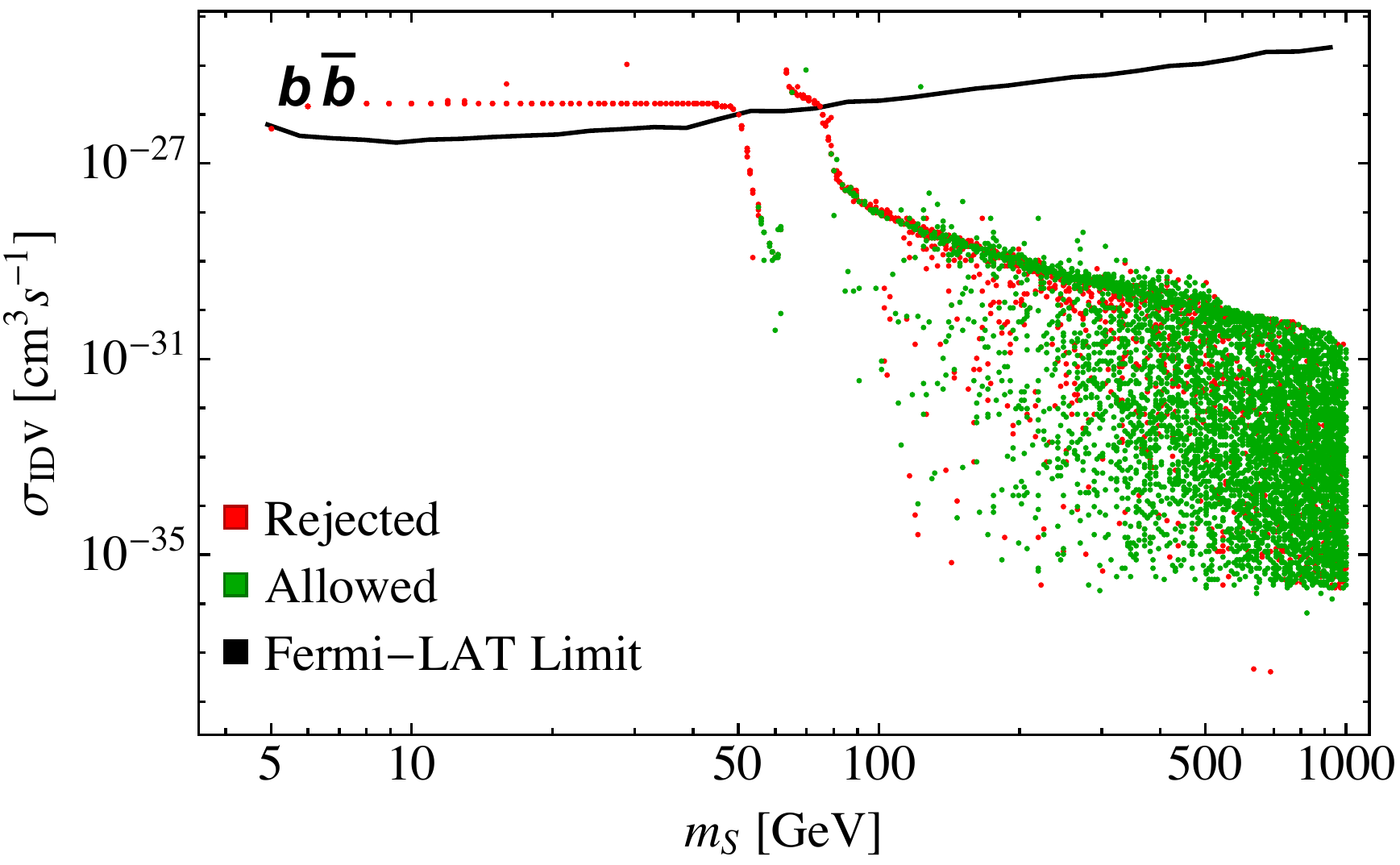}
$\;$
\includegraphics[width=0.5\textwidth]{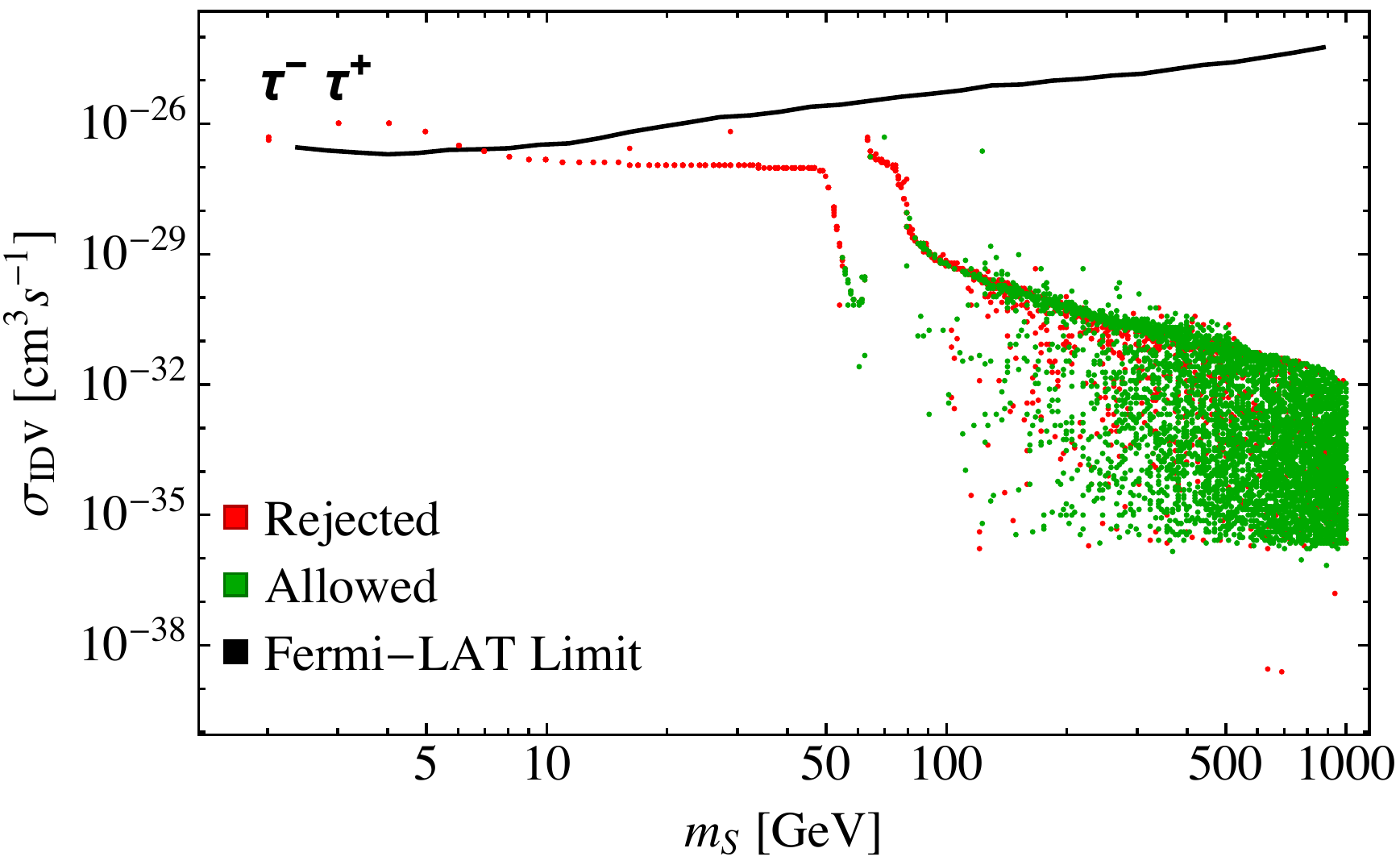}
$\;$
\includegraphics[width=0.5\textwidth]{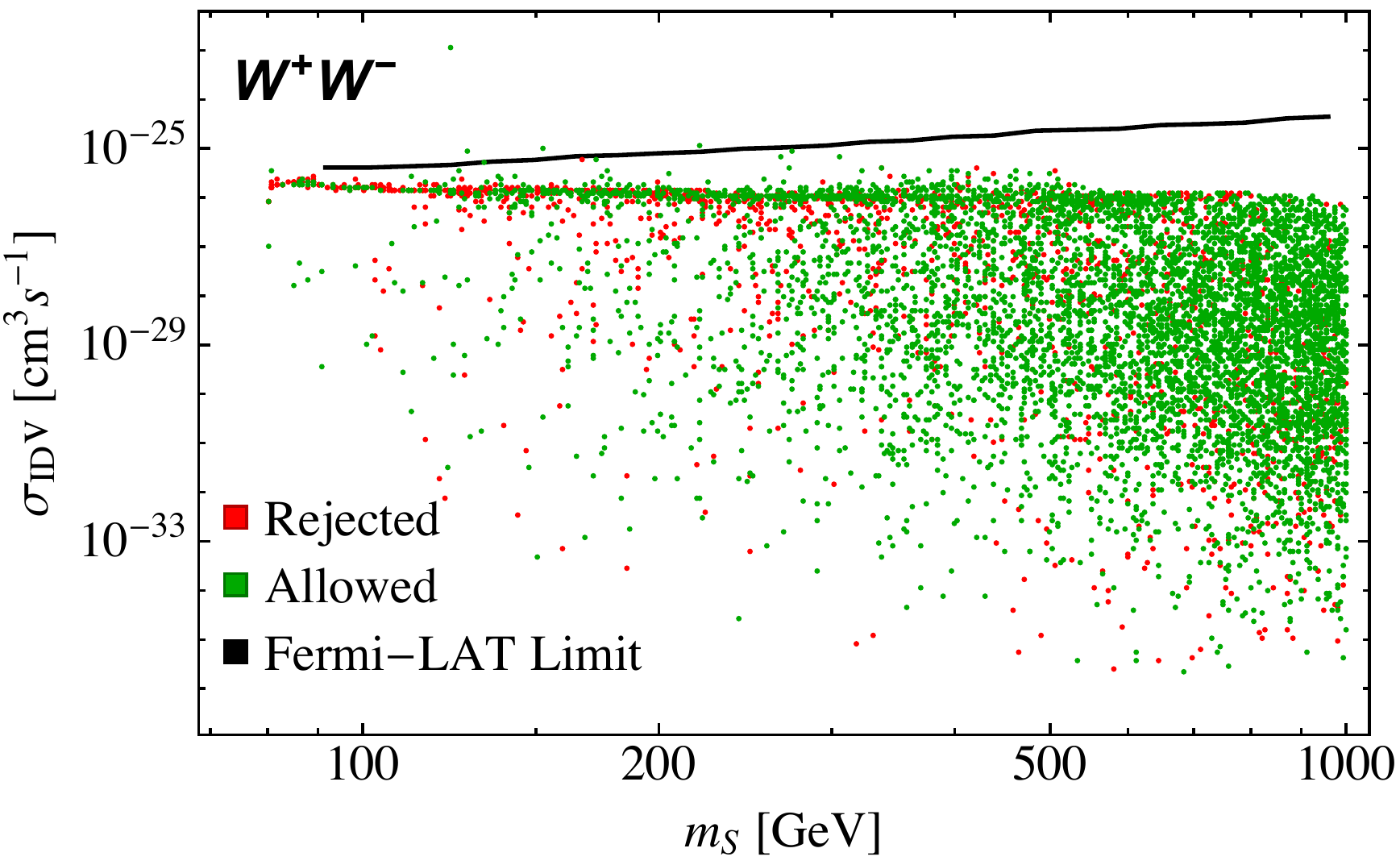}
\caption{Product of dark matter annihilation cross section and velocity at $v\approx 0$
as a function of the mass of the dark matter 
candidate $S$ for the $b\bar{b}$, $\tau^+\tau^-$, and $W^+W^-$ final states 
for the theoretically available points in our scan. Points 
labeled as ``rejected" are points that do not satisfy at least one of the invisible width, 
Higgs signal strength, or self-interaction constraints.}
\label{fig:IDbb}
\end{figure}

In Fig.~\ref{fig:IDbb},
we show our results along with the Fermi-LAT limits for the
$b\bar{b}$, $\tau^+ \tau^-$, and $W^+ W^-$ final states. 
Because $\sigma_\text{ID} v$ is evaluated at threshold, the lower bound
is dictated by the kinematic threshold and each plot has a different lower bound. Below 
$m_S \approx m_{h_1}/2$, the cross sections are relatively insensitive to $m_S$. 
In this region, the $h_{1,2}SS$ 
vertices are proportional to $\lambda_{\varphi\varphi SS} w$, which appears in both 
$\sigma_\text{ID}$ and $\langle \sigma v \rangle$ (which feeds into the relic abundance via the
Boltzmann equation~\cite{Kolb:1990vq}).  As a consequence, any change in $w$ leads 
to a corresponding change in the value for $\lambda_{\varphi\varphi SS}$ to give
the correct relic abundance so that the product
$\lambda_{\varphi\varphi SS} w$ remains constant for a given value of $m_S$. 
In any case, the points for  $m_S \lesssim m_{h_1}/2$
are almost always ruled out by $\text{BR}_\text{inv}$ when the decay $h_1 \to SS$ is
kinematically allowed because of the resulting large $\Gamma_\text{inv}$.
The dip at $m_S \approx m_{h_1}/2$ is due to the Higgs resonance in the annihilation cross section 
entering in the calculation of the relic abundance, 
forcing $\lambda_{\varphi\varphi SS}$ to be small to give the correct relic abundance 
and resulting in a dip in the indirect detection cross section. 

While  the indirect detection 
limits do reject some parameter points for $b\bar{b}$ and $\tau^-\tau^+$ final states, 
most of these were already rejected by 
previous constraints. Only a few points are rejected for the $W^+W^-$ final state, but modest 
improvements in experimental sensitivity will start ruling out regions of the parameter space allowed
by other constraints.

\subsection{Constraints from Direct Detection}

Now that all the theoretical constraints and various experimental constraints have been
applied to the parameter scan, we turn to our original purpose of confronting the surviving
points with the direct detection experimental limits.
In this section  we compare our parameter points to the limits from the XENON1T experiment~\cite{Aprile:2017iyp}. 
We want to see if patterns
emerge with respect to regions in the parameter space where the
 direct detection cross section is suppressed. We start with an overview of the direct
detection cross sections ($\sigma_\text{DD}$) for the scan of parameter points and then examine 
specific characteristics of the results.

\begin{figure}
\centering
\includegraphics[width=0.48\textwidth, trim= 0 150 0 150]{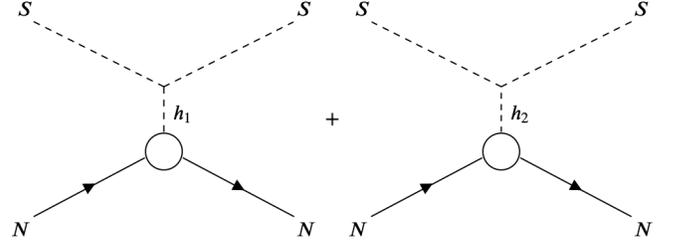}
\caption{Feynman diagrams for the $t$-channel exchange involved in direct detection, where $N$ is the nucleon.}
\label{fig:feynDD}
\end{figure}

In our model, the Higgs boson $t$-channel exchange from a Higgs portal is replaced with $t$-channel exchange
of the $h_1$ and $h_2$ which is shown in Fig.~\ref{fig:feynDD}. 
The direct detection cross section for scalar DM with a Higgs
portal is given by~\cite{Cline:2013gha}
\begin{equation}
\sigma_\text{DD} = {1\over {4\pi}} {{M_N^2}\over{(m_S+M_N)^2}} {{ f_N^2 M_N^2}\over{v^2}} 
\left( {\lambda_{hs}\over{ m_h^2}}\right)^2  \label{fig:ddcs}
\end{equation}
with $m_h$ the Higgs boson mass,  $\lambda_{hs}$ the Higgs-scalar DM coupling,  
$M_N=938.95$~MeV the nucleon mass, and $f_N=0.30$ the Higgs nucleon coupling
so that $h_1$ and $h_2$ exchange results in the following substitution
\begin{eqnarray}
\left({\lambda_{hs}\over{ m_h^2}}\right)^2 &  \to &
\left(  { {g_{{h_1} SS}\cos\alpha}\over{m_{h_1}^2}} + { {g_{{h_2}SS}\sin\alpha }\over{m_{h_1}^2}} \right)^2  \\
& = & 4 \cos^2\alpha \sin^2 \alpha \lambda_{\varphi\varphi SS}^2 w^2 \left(  { {1}\over{m_{h_1}^2}} - {{1}\over{m_{h_1}^2}} \right)^2,
\nonumber
\end{eqnarray}
where we used the relations from Eqns.~\ref{eqn:ghss} and \ref{eqn:gpss}.  One notes the cancellation between
the two $t$-channel exchanges  and, more importantly, that the direct 
detection cross section is proportional
to $\lambda_{\varphi\varphi SS}^2$ which, as pointed out above, is fitted to give the correct relic abundance.

\begin{figure}
\centering
\includegraphics[width=0.48\textwidth]{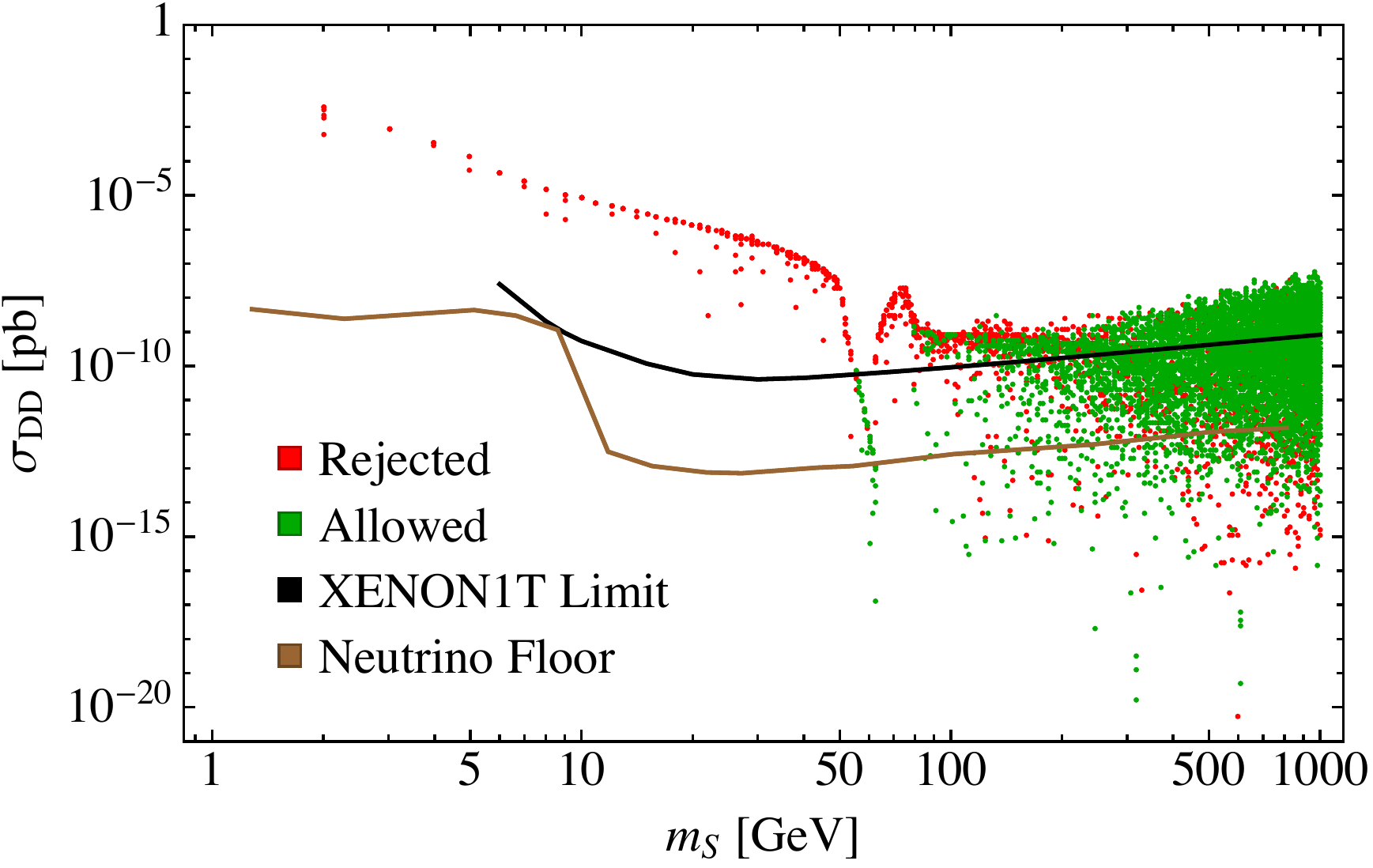}
\caption{The dark matter direct detection cross section as a function of $m_S$ 
for the 8148 theoretically allowed points from our scan of 10,000 points. Points labeled as 
``rejected" are points that do not satisfy at least one of the invisible width, Higgs 
signal strength, self-interaction, or indirect detection constraints.}
\label{fig:ddplot}
\end{figure}

Fig.~\ref{fig:ddplot} shows the direct detection cross sections
 calculated using micrOMEGAs~\cite{Belanger:2018ccd}
for the 8148 points of the original 10,000 points
that passed the theoretical constraints in our parameter scan. 
The red points were rejected by at least one of the invisible width, Higgs signal strength,
dark matter self-interaction, or indirect detection constraints. We remind the reader that, for a given value
of $m_S$, we vary $m_{h_2}$, $\cos\alpha$, and $w$. We fit $\lambda_{\varphi\varphi SS}$ to give
the correct relic abundance, and since $\lambda_S$ is mainly constrained by the self-interaction 
cross section, we chose a value that passes this constraint. 

In the region below $m_S\approx m_{h_1}/2$, $\sigma_\text{DD}$ is
mainly determined by $m_S$ as can be seen from Eqn.~\ref{fig:ddcs} with slight variations due 
to the value of $\alpha$, 
and is largely independent of the other parameters. 
This is for the same reason as with indirect detection 
as discussed in section~\ref{sec:ID}: the 
$h_{1,2}SS$ vertices are proportional to $\lambda_{\varphi\varphi SS} w$, which appears in both 
$\sigma_\text{DD}$ and $\langle \sigma v \rangle$, so that any change in $w$ leads 
to a corresponding change in the value for $\lambda_{\varphi\varphi SS}$ to give
the correct relic abundance and the product
$\lambda_{\varphi\varphi SS} w$ remains constant for a given value of $m_S$.
Likewise,  as also discussed in section~\ref{sec:ID}, 
the dip in $\sigma_\text{DD}$ around $m_S \approx m_{h_1}/2 $ is due to the Higgs resonance
where $\lambda_{\varphi\varphi SS}$ needs to be small to compensate for the enhancement
in the $SS$ annihilation cross section to obtain the correct relic abundance.

\begin{figure}[t]
\centering
\includegraphics[width=0.48\textwidth]{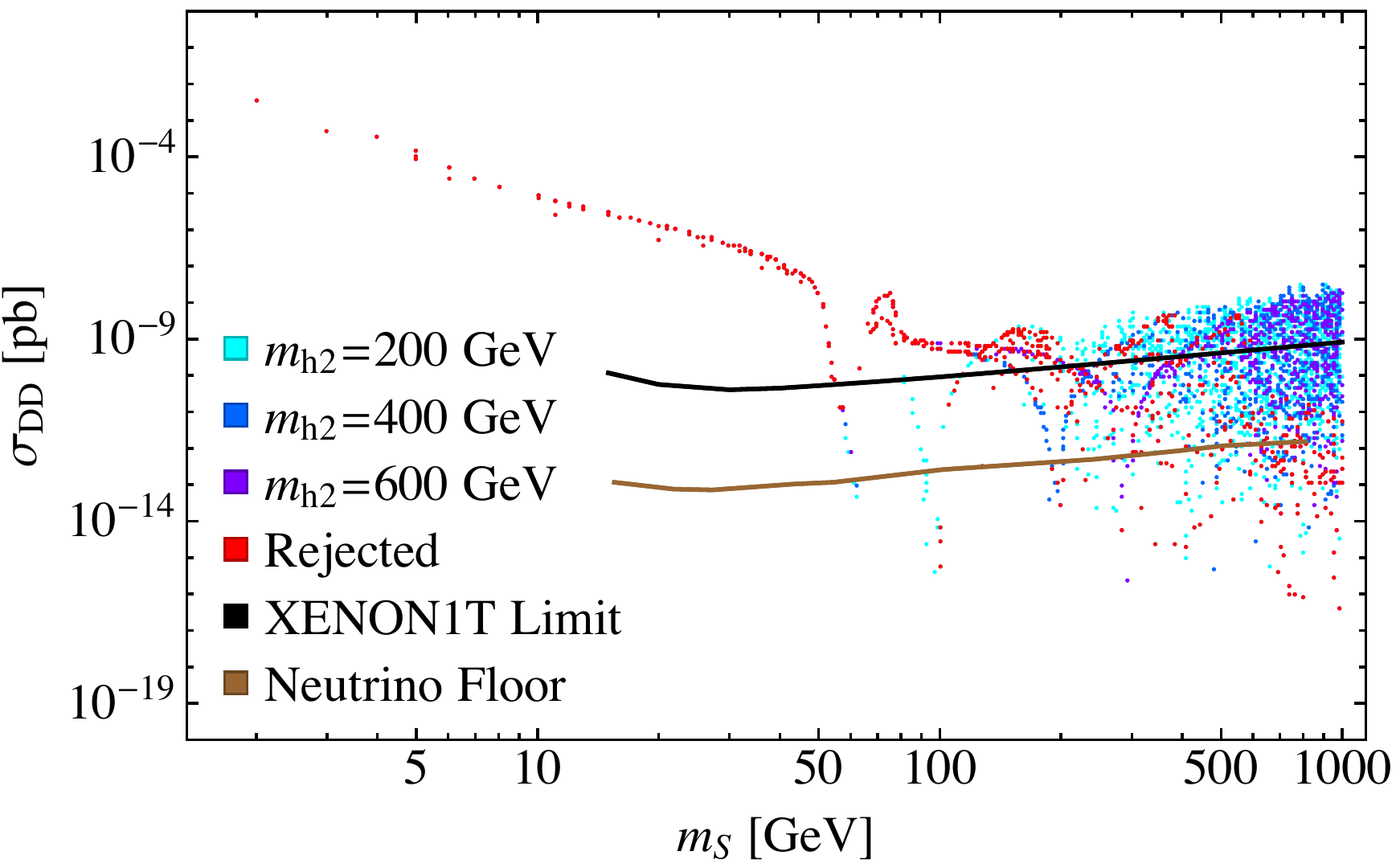}
\caption{The dark matter direct detection cross section as a function of $m_S$ 
for random theoretically allowed points with fixed 
values of $m_{h_2}$. Points labeled as ``rejected" are points that do not satisfy at 
least one of the invisible width, Higgs signal strength, self-interaction, or 
indirect detection constraints.}
\label{fig:mpres}
\end{figure}

\begin{figure}
\centering
\includegraphics[width=0.48\textwidth]{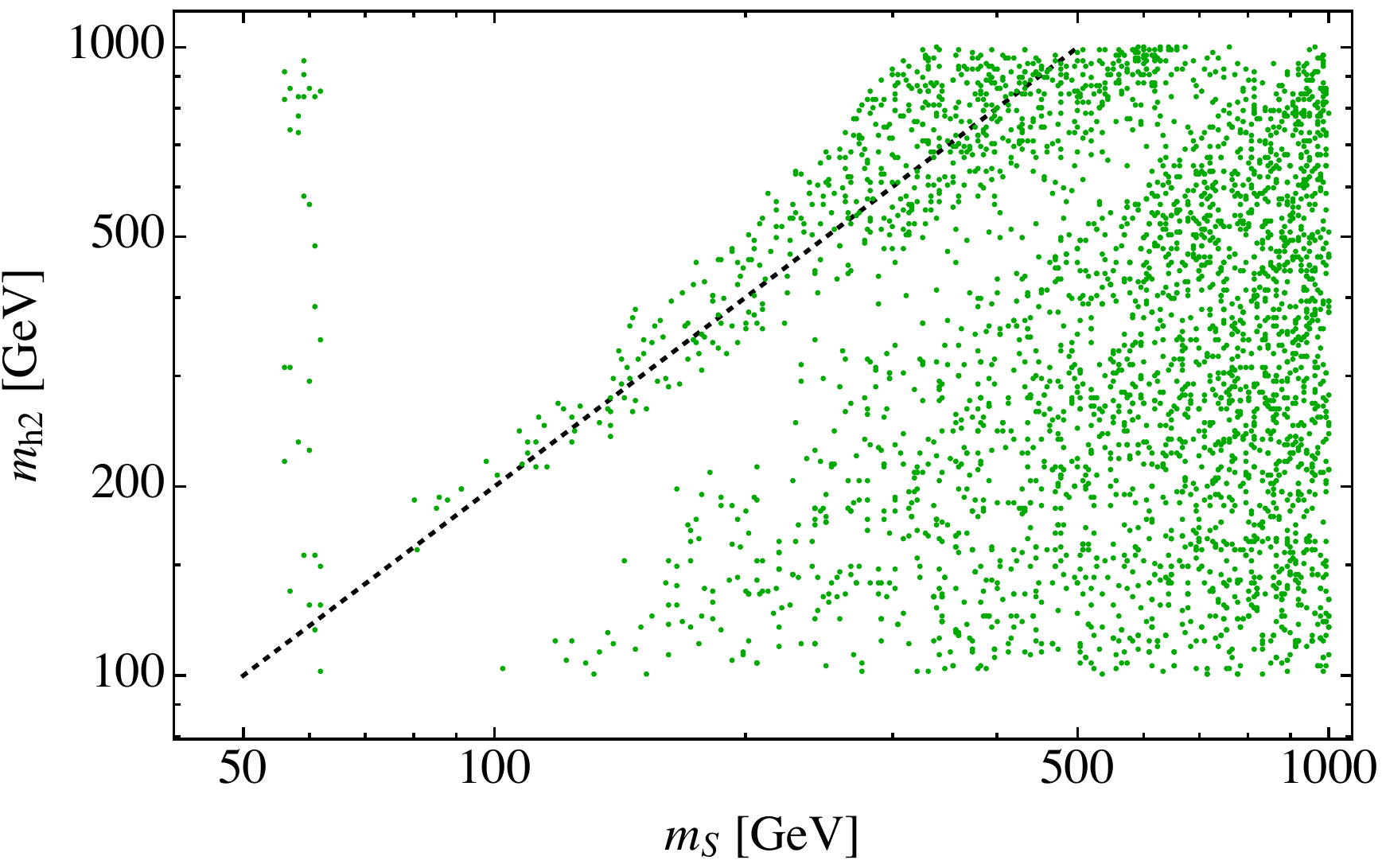}
\caption{All points allowed by invisible width, Higgs signal strength, self-interaction, 
indirect detection, and direct detection constraints plotted on the $m_{h_2}$-$m_S$ plane. 
The dotted line indicates $m_{h_2} = 2 m_S$.
}
\label{fig:mscallin}
\end{figure}

The region for $m_S \gtrsim 80$~GeV shows numerous parameter points not ruled out by
direct detection limits. There are two effects contributing to this. The first is due to the
resonance effect of the portal scalar when $m_S \approx m_{h_2}/2$, which is analogous to the 
Higgs resonance effect described
above~\cite{LopezHonorez:2012kv,Arcadi:2017kky,Azevedo:2018oxv}. 
Near the $h_2$ resonance, the $SS$ annihilation cross section increases, 
requiring a smaller value for $\lambda_{\varphi\varphi SS}$ to obtain the correct relic
abundance, resulting in a small direct detection cross section. 
This is illustrated in Fig.~\ref{fig:mpres} which shows, in addition to the Higgs/$h_1$
resonance, dips in the direct detection cross section at $m_S=100$, 200 and 300~GeV
corresponding to $m_{h_2}=200$, 400 and 600~GeV respectively. 
The linear relationship corresponding to $m_{h_2} \approx 2 m_S$ shows up clearly as a cluster of points along the diagonal in Fig.~\ref{fig:mscallin}, 
which plots the 
parameter points allowed by direct detection on a plot of $m_{h_2}$ vs $m_S$. The cluster
of points in the vertical band at $m_S \approx 62.5$~GeV corresponds to the Higgs resonance,
and the cluster of points below the diagonal in the bottom right portion of the
plot reflects a second effect which we discuss next. The lack of points 
along $m_S=m_{h_2}$ simply  reflects the fact that there are no similar effects in that region.

\begin{figure}[t]
\centering
\includegraphics[width=0.48\textwidth]{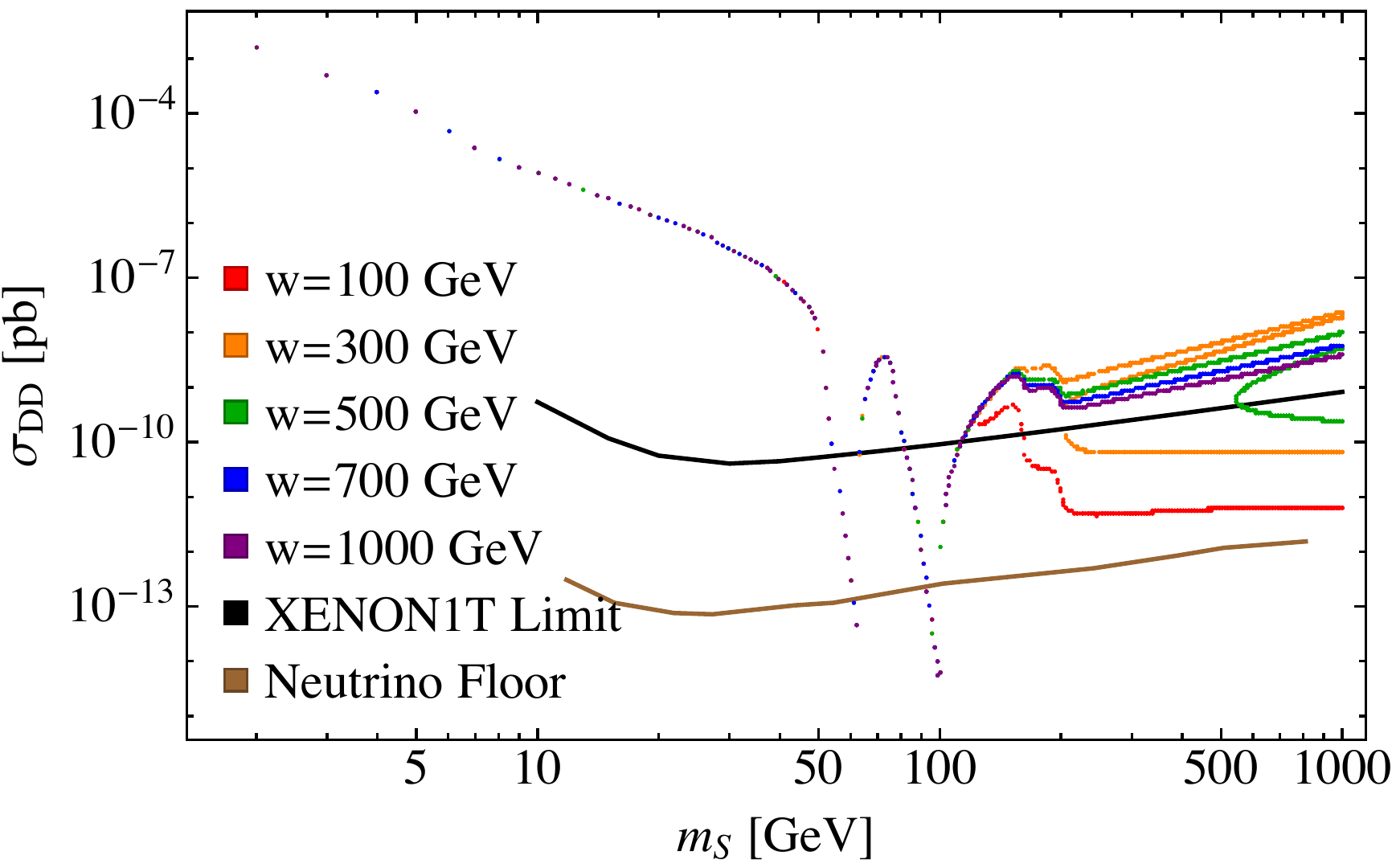}
\caption{The dark matter direct detection cross section as a function of $m_S$ 
 for theoretically allowed points with 
$\alpha=0.2$, $m_{h_2}=200~\text{GeV}$, and the fixed values of $w$ given in the legend.
}
\label{fig:duckbeak}
\end{figure}

This second effect results in a big spread of the direct detection cross section 
and the allowed parameter points, and is more interesting due 
to non-trivial relationships between the parameters $w$ and $\lambda_{\varphi\varphi SS}$ 
and how this influences
the annihilation cross sections as described in Section~\ref{sec:scan}. We 
refer to Fig.~\ref{fig:duckbeak} to examine the details of this behavior. 
For $m_S < m_{h_i}$ where $i=1$ or 2, the annihilation cross section is dominated
by $SS \to W^+W^-$ and $ZZ$, while for 
$m_S > m_{h_i}$ the annihilation cross sections into $h_1$ and $h_2$ 
become important for achieving the correct relic abundance. 
In Fig.~\ref{fig:duckbeak}, we see that the resulting
direct detection cross section drops at $m_S=m_H=125$~GeV and again at $m_S=m_{h_2}=200$~GeV,
the value used for $m_{h_2}$ in this figure. These points 
correspond to where the $SS\to h_i h_i $ annihilation channels
open up so that a smaller value of $\lambda_{\varphi\varphi SS}$ is needed to achieve the correct
relic abundance.

When $m_S > m_{h_i}$, 
the direct detection cross section in Fig.~\ref{fig:duckbeak} depends
on the value of $w$ because it is the product
$w \lambda_{\varphi\varphi SS}$ that enters the
 expressions for the $s$-channel 
annihilation cross sections for $SS \to h_i h_j$, where $i,j=1$ or 2. 
In this situation, as seen in Fig.~\ref{fig:lamOmega}, 
there can be multiple values of $\lambda_{\varphi\varphi SS}$ 
that give the correct relic abundance 
for a given set of the free parameters, $m_{h_2}$, $m_S$, $\alpha$, and $w$,
due to the peak in $\Omega_{DM}$ at $\lambda_{\varphi\varphi SS} \approx m_s^2/2w^2$.
This results in the multiple values for the direct detection cross section seen in Fig.~\ref{fig:duckbeak}. 
Referring to Fig.~\ref{fig:lamOmega}, we can see that this situation only arises for intermediate values of $w$. 
This is because for small values of $w$ the peak shifts to large values of $\lambda_{\varphi\varphi SS}$
where $\Omega_{DM}$ falls below the observed value,
while for large values of $w$ the calculated value of $\Omega_{DM}$
sits above the measured value until after the peak.
As such, for small and large values of $w$, there is
only one solution for $\lambda_{\varphi\varphi SS}$. 
The multiple values of $\lambda_{\varphi\varphi SS}$ for intermediate values of $w$ 
result in multiple values for the direct detection cross section, 
although it should be noted that the additional points with large values 
of $\lambda_{\varphi\varphi SS}$ are more likely to be inconsistent with direct detection limits.

\begin{figure}[t]
\centering
\includegraphics[width=0.48\textwidth]{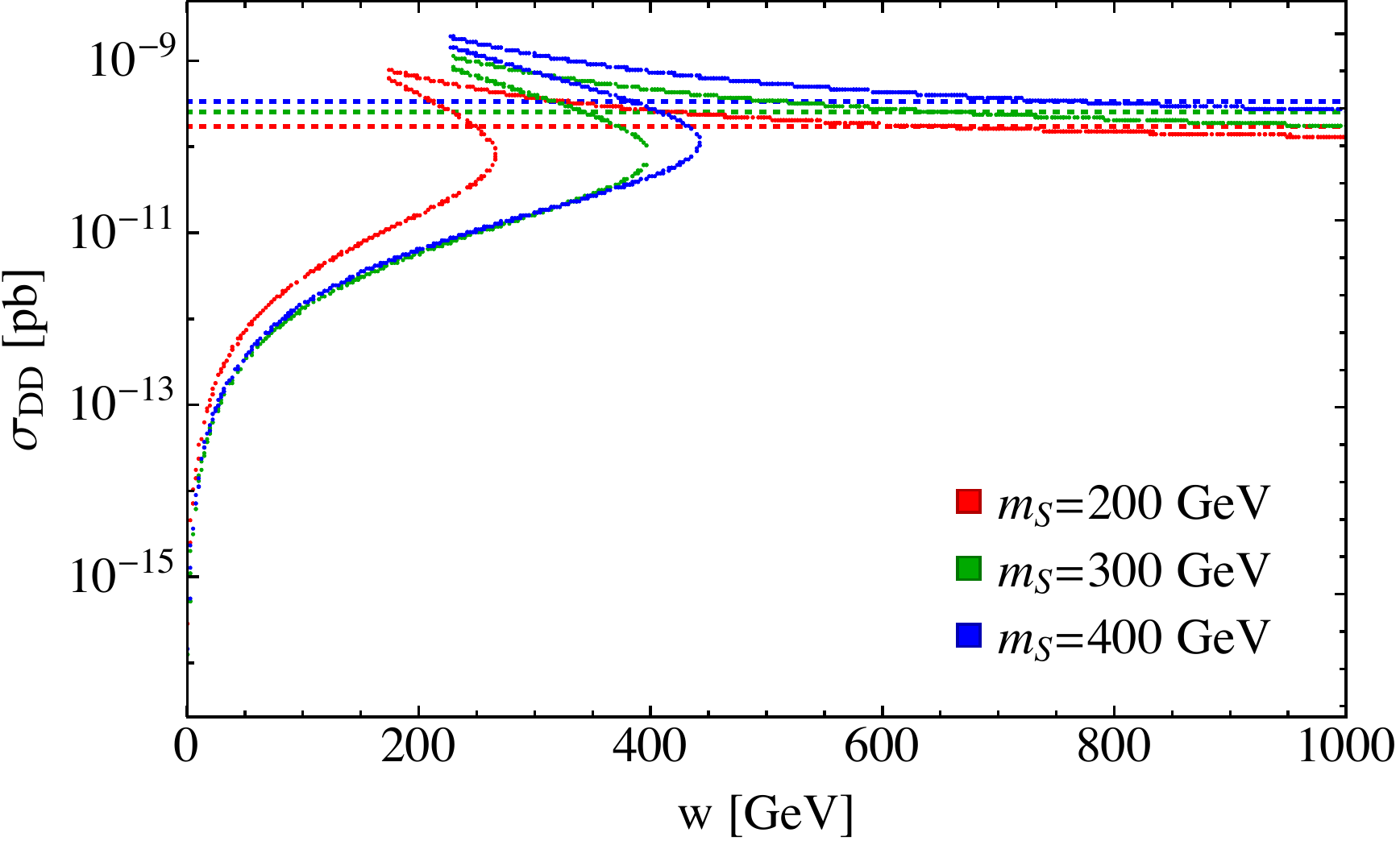}
\caption{Dark matter direct detection cross section as a function of $w$ for 
theoretically available points with $\alpha=0.1$, $m_{h_2}=200~\text{GeV}$, 
and the fixed values of $m_S$ given in the legend. The dotted lines correspond to the XENON1T limit for 
each corresponding value of $m_S$. For each value of $m_S$ shown, 
there is a intermediate range of $w$ values that have no points below the direct detection 
cross section limit.}
\label{fig:icepick}
\end{figure}

We can see how the solutions evolve with $w$ 
from a different perspective in Fig.~\ref{fig:icepick}, where we plot 
$\sigma_{DD}$ versus $w$ while keeping the other parameters fixed and as usual fitting 
$\lambda_{\varphi\varphi SS}$ to give the correct relic abundance. The horizontal
lines are the XENON1T limits, so points below the lines are allowed and points above are ruled out.
The regions of parameter space at both small and large values of $w$ are allowed by the direct detection limits.
In the intermediate region, starting with small values of $w$, 
there are multiple values for the direct detection cross sections reflecting the multiple
solutions for $\lambda_{\varphi\varphi SS}$ that give the correct relic abundance.
In this region, some solutions give rise to large direct detection cross sections that are ruled out by 
experimental limits while others are allowed. As $w$ increases further, we leave the region 
of multiple solutions and the remaining solutions are ruled out by direct detection limits until eventually 
they fall below the XENON1T limits.
The size of the ruled out region depends on the cancellations of the dark matter
annihilation cross sections for the available scalar channels. In our model,
when kinematically allowed, the $SS\rightarrow h_2 h_2$ channel dominates. However,
this region could be larger for cases where multiple scalar channels 
are comparable in importance.

\subsection{Effect of taking $\lambda_{HHSS} \neq 0$ }
\label{sec:lnez}

\begin{figure}[t]
\centering
\includegraphics[width=0.48\textwidth]{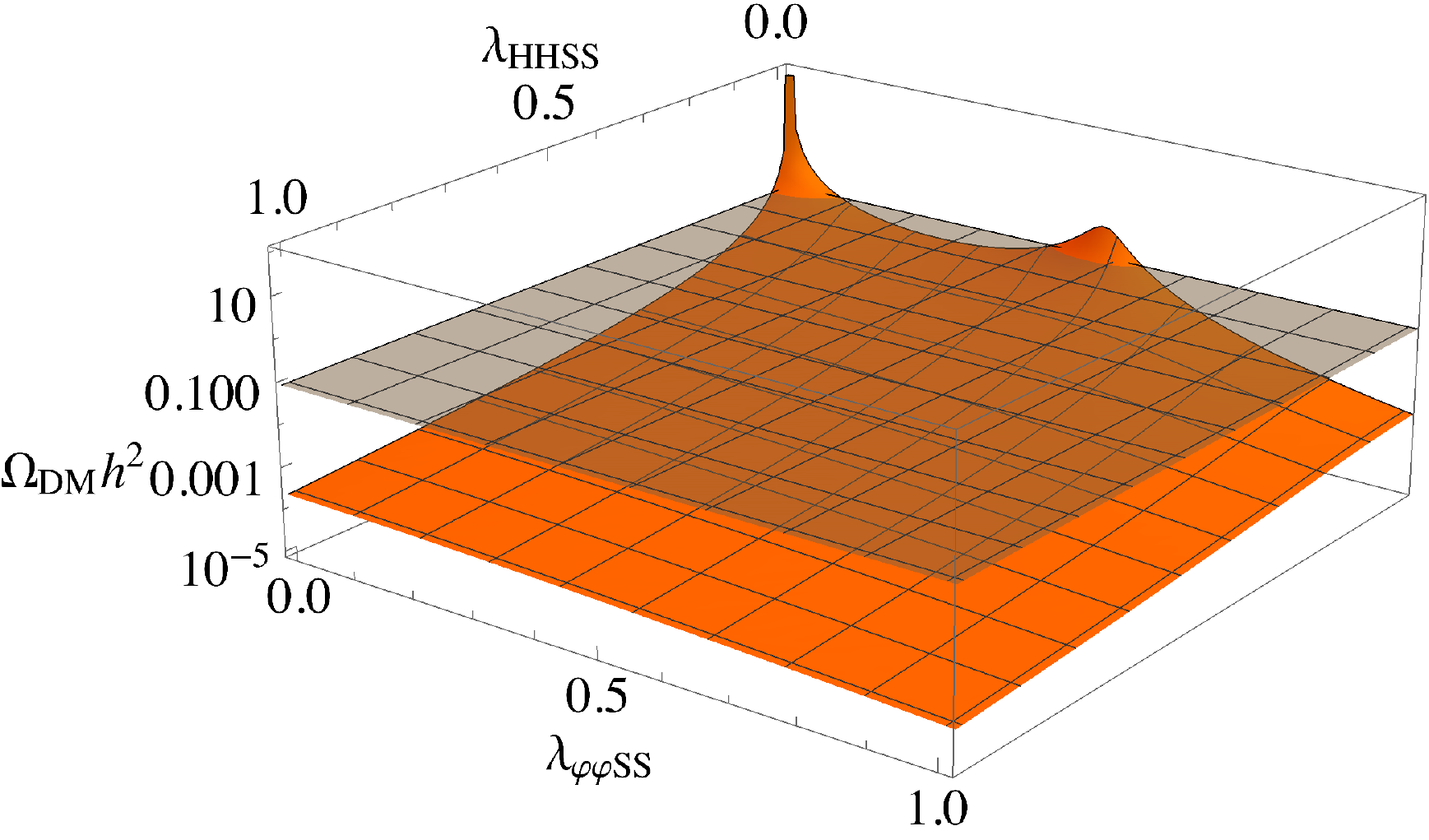}
\caption{ Dark matter relic abundance as a function of 
$\lambda_{\varphi\varphi SS}$ and $\lambda_{HHSS}$ for $\alpha=0.2$, $m_{h_2}=200~\text{GeV}$, $m_S=300~\text{GeV}$, 
and $w=300~\text{GeV}$. The grey plane is for the measured value
of $\Omega_\text{DM} =0.1200(12)\; h^{-2}$~\cite{Tanabashi:2018oca}.}
\label{fig:lamlam}
\end{figure}

We end this section with some comments on the consequences of not setting $\lambda_{HHSS}$ 
to zero in Eqn.~\ref{eqn:V}. We chose $\lambda_{HHSS} = 0$ to highlight the interplay between parameters of the model,
and altering this choice will not affect our conclusions. 
Allowing $\lambda_{HHSS} \neq 0$ introduces an additional parameter 
so that for this case it is
 a linear combination of $\lambda_{\varphi\varphi SS}$ and $\lambda_{HHSS}$
 that is fitted to reproduce the observed relic abundance.
This gives a family of solutions for these two 
Lagrangian parameters when keeping the rest of the parameters fixed. This is illustrated in 
Fig.~\ref{fig:lamlam} where the relic abundance is plotted as a function of $\lambda_{\varphi\varphi SS}$ 
and $\lambda_{HHSS}$ with the other parameters fixed to 
the same values given in Fig.~\ref{fig:lamOmega}; 
$\alpha=0.2$, $m_{h_2}=200$~GeV, $m_S=300$~GeV, with the choice $w=300$~GeV.
We see that there is now a continuum
of solutions, with our choice in this paper corresponding to solutions
where $\lambda_{HHSS} = 0$. 
Rotating from the $\lambda_{HHSS} = 0$ axis to the
$\lambda_{\varphi\varphi SS}=0$ axis simply corresponds to another choice of parameters.
For the parameters used in Fig.~\ref{fig:lamlam}, 
we can obtain the correct relic abundance for a continuum of $\lambda_{HHSS}$
and $\lambda_{\varphi\varphi SS}$ values,
but the multi-valueness of $\Omega_{DM}$ 
 is only present near the $\lambda_{HHSS} = 0$ axis,
so that when $\lambda_{HHSS}\neq 0$ the prediction for the direct detection cross section 
is more straightforward.
Thus, taking $\lambda_{HHSS} \neq 0$ does not qualitatively change
our results but misses the subtleties and richness of 
the effects that are discussed in this paper. 

\section{Conclusions}
\label{sec:conclusions}

We studied a simple model of scalar DM with a scalar portal that can mix with the SM Higgs.
Our purpose was to explore 
regions of parameter space with a suppressed 
direct detection cross section for a Higgs portal
model. We found that 
even in this simple model there remains significant regions of parameter space that are not ruled
out by direct detection measurements, with many points lying below the neutrino floor. Three
of the mechanisms leading to these regions have been discussed previously;
a small Higgs-portal mixing angle leading to a small coupling with the DM, 
the Higgs resonance effect which requires a small DM-portal coupling to compensate for the 
enhanced DM annihilation cross section due to the Higgs resonance, and the similar effect
as a result of the portal resonance.

An additional effect is the result of a heavy DM particle
with a lighter portal. This opens up new DM annihilation channels so that the parameters controlling
this annihilation need to compensate, resulting in a smaller direct detection cross section.
For certain regions of the parameter space, 
destructive interference between diagrams leads to multiple solutions for the DM-portal couplings, resulting in
a spread of allowed parameter points. We therefore find, contrary to common lore, 
that even in a very simple model of DM there
are sizeable regions of parameter space that are still allowed by direct detection limits.

\begin{acknowledgements}
The authors thank Thomas Gregoire, Heather Logan, Alex Poulin and Daniel Stolarski for helpful conversations.
This work was supported by the Natural Sciences and Engineering Research Council of Canada
under grant number SAPIN-2016-00041. 
\end{acknowledgements}

\begin{appendices}

\section{Interference in the dark matter annihilation amplitude}
\label{appendix}

\begin{figure}
\centering
%\begin{subfigure}{.25\textwidth}
  \centering
  \includegraphics[width=.3\linewidth, trim= 100 250 100 250]{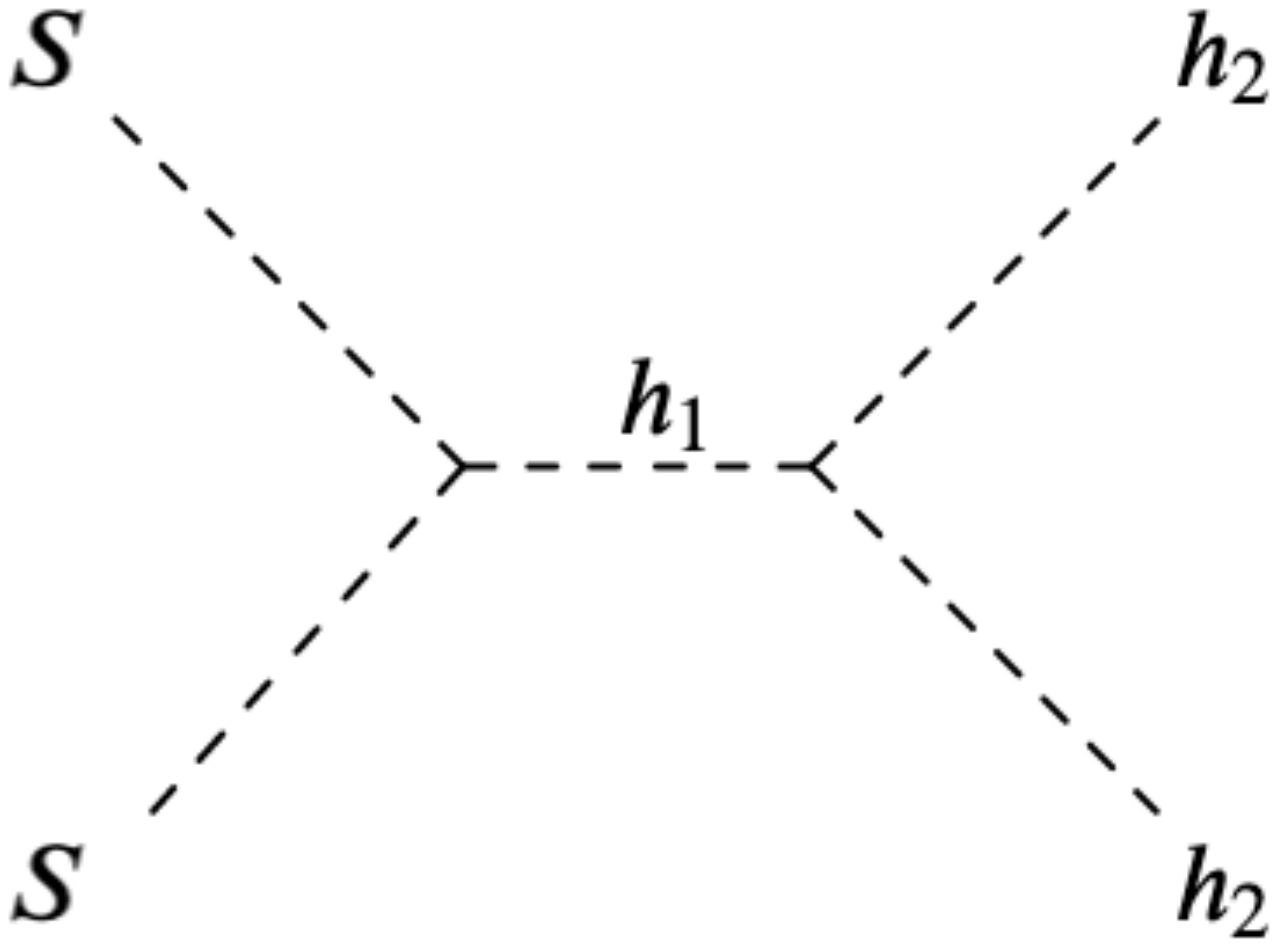}
%  \caption{}
%\end{subfigure}%
%\begin{subfigure}{.25\textwidth}
  \centering
  \includegraphics[width=.3\linewidth, trim= 100 250 100 250]{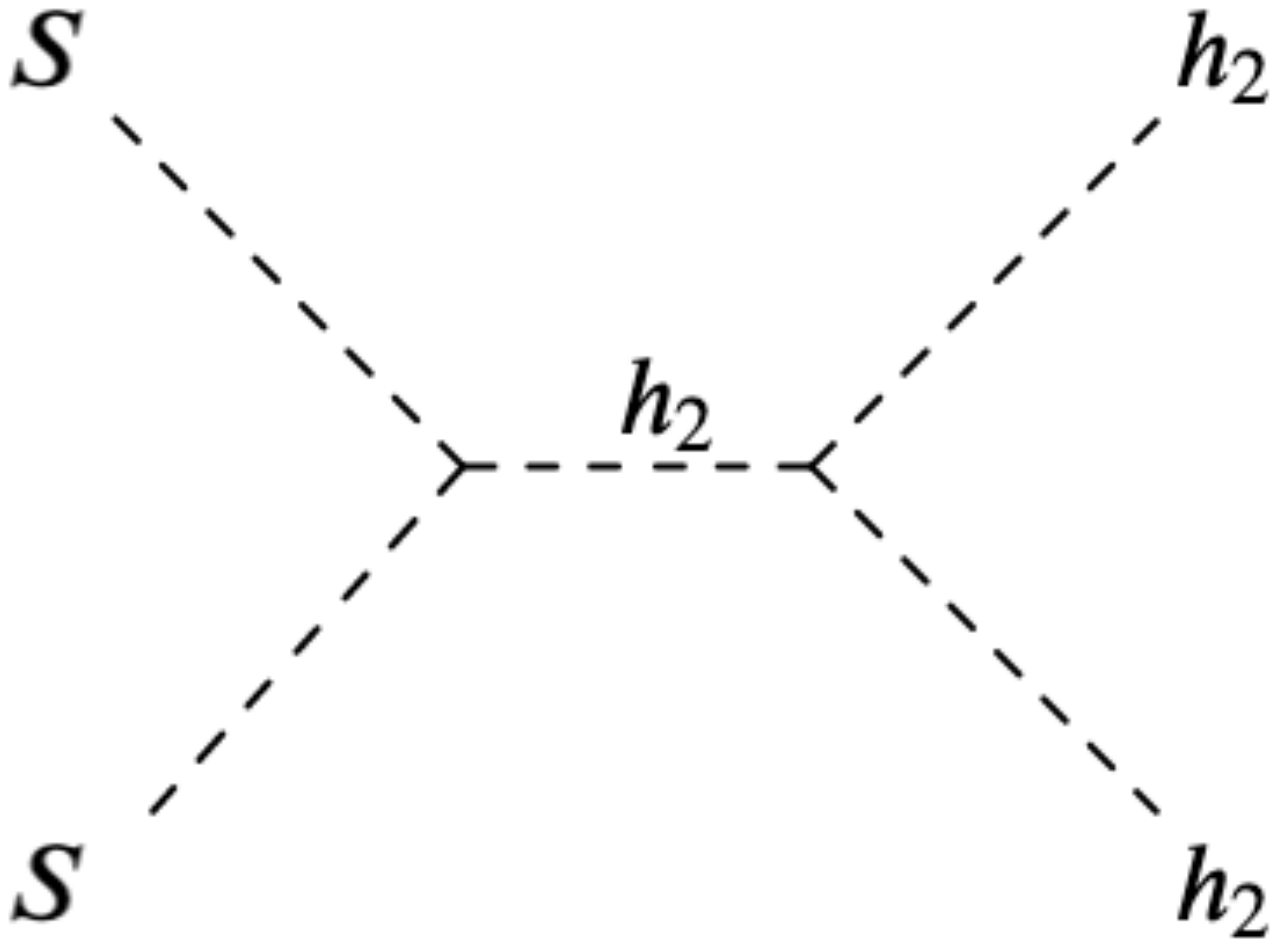}
%  \caption{}
%\end{subfigure}\\
%\begin{subfigure}{.25\textwidth}
  \centering
  \includegraphics[width=.3\linewidth, trim= 100 250 100 250]{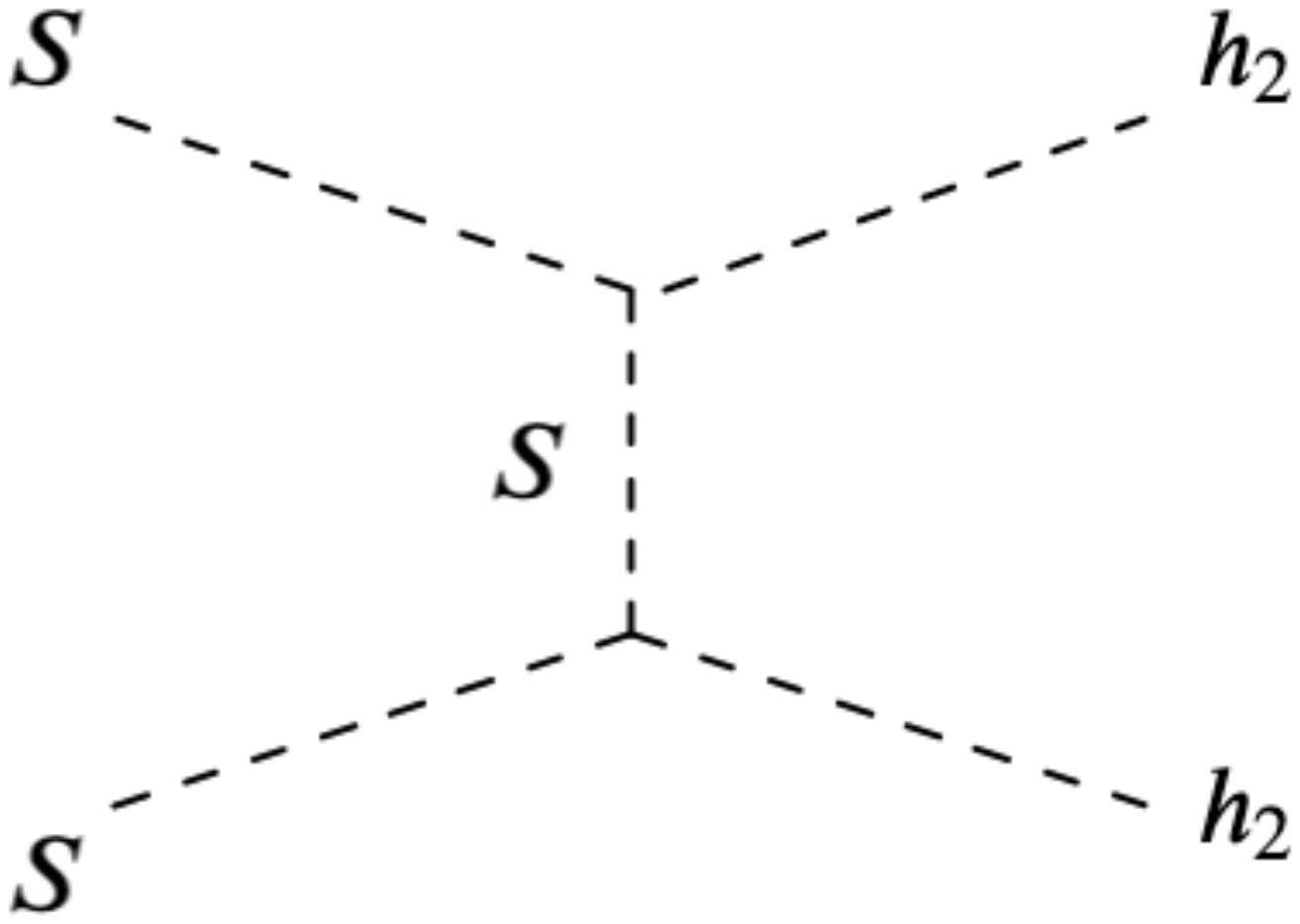}
%  \caption{}
%\end{subfigure}%
%\begin{subfigure}{.25\textwidth}
  \centering
  \includegraphics[width=.3\linewidth, trim= 100 250 100 250]{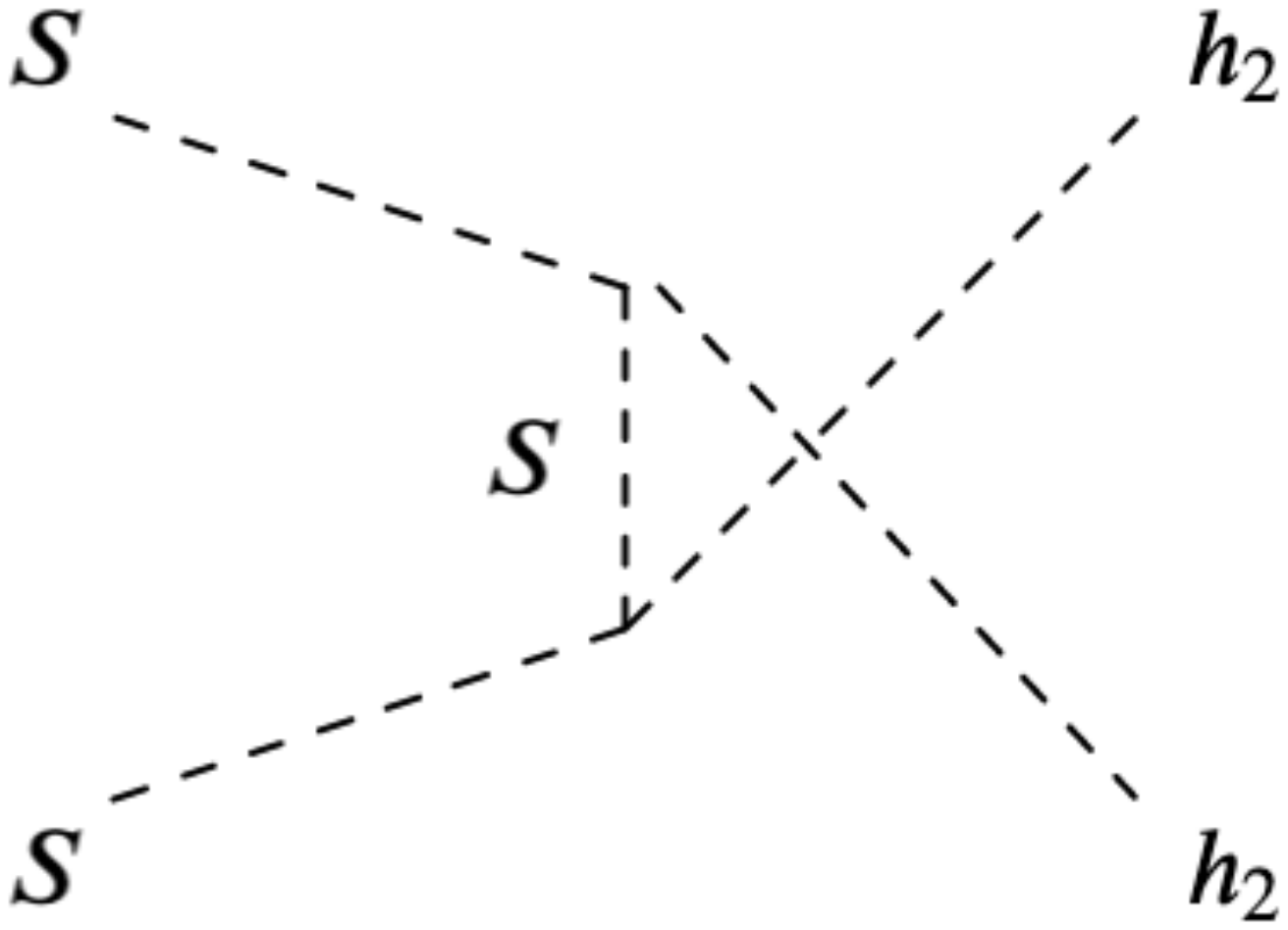}
%  \caption{}
%\end{subfigure}\\
%\begin{subfigure}{.25\textwidth}
  \centering
  \includegraphics[width=.3\linewidth, trim= 100 250 100 250]{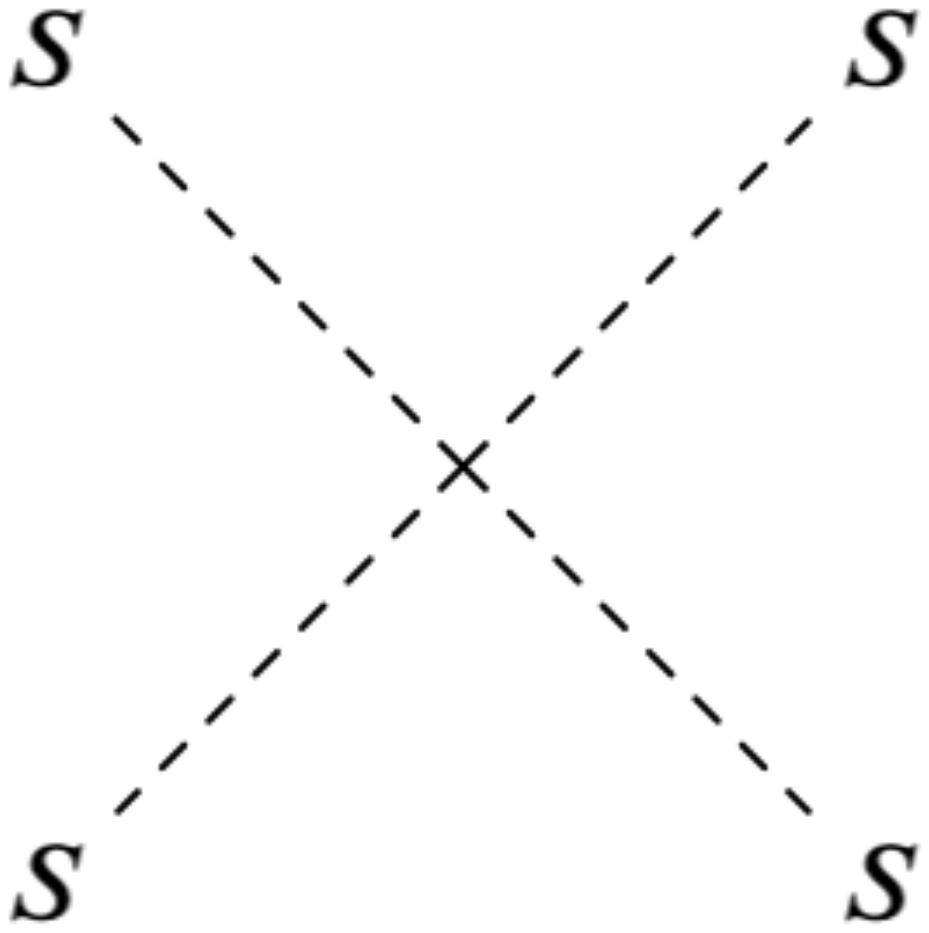}
%  \caption{}
%\end{subfigure}\\
\caption{Feynman diagrams contributing to the $SS\rightarrow h_2h_2$ process.}
\label{fig:feynman}
\end{figure}

The model presented in section~\ref{sec:model} features a dark matter candidate $S$ which 
couples only to the other two physical scalars $h_1$ and $h_2$, giving three annihilation 
channels: $SS\rightarrow h_1h_1$, $SS\rightarrow h_1h_2$, and $SS\rightarrow h_2h_2$. 
At tree-level, each of these channels features the five diagrams shown in Fig.~\ref{fig:feynman}: 
two $s$-channels with
 $h_1$ and $h_2$ mediators, $t$- and $u$-channels with an $S$ mediator, and a quartic 
 vertex.

The relevant vertices are given by
\begin{eqnarray}
g_{h_1h_1h_1} & = & 6i(\lambda_H v c_\alpha^3 - \lambda_4 w c_\alpha^2 s_\alpha + \lambda_4 v c_\alpha s_\alpha^2 \nonumber\\
&&	 - \lambda_\varphi w s_\alpha^3) \\
g_{h_1h_1h_2} & = & 2i(\lambda_4 w c_\alpha^3 - \left(2\lambda_4 -3\lambda_H\right) v c_\alpha^2 s_\alpha \nonumber\\
&&- \left(2\lambda_4 -3\lambda_\varphi\right) w c_\alpha s_\alpha^2 + \lambda_4 v s_\alpha^3)\\
g_{h_1h_2h_2} & = & 2i(\lambda_4 v c_\alpha^3 + \left(2\lambda_4 -3\lambda_\varphi\right) w c_\alpha^2 s_\alpha  \nonumber\\
&&- \left(2\lambda_4 -3\lambda_h\right) v c_\alpha s_\alpha^2 - \lambda_4 w s_\alpha^3) \\
g_{h_2h_2h_2} & = & 6i(\lambda_\varphi w c_\alpha^3 + \lambda_4 v c_\alpha^2 s_\alpha + \lambda_4 w c_\alpha s_\alpha^2 \nonumber\\
&&	 + \lambda_H v s_\alpha^3) \\
g_{h_1SS} & = & - 2i \lambda_{\varphi\varphi SS} w s_\alpha \label{eqn:ghss}\\
g_{h_2SS} & = & 2i \lambda_{\varphi\varphi SS} w c_\alpha  \label{eqn:gpss}\\
g_{h_1h_1SS} & = & 2i \lambda_{\varphi\varphi SS} s_\alpha^2\\
g_{h_1h_2SS} & = & -2i \lambda_{\varphi\varphi SS} c_\alpha s_\alpha\\
g_{h_2h_2SS} & = & 2i \lambda_{\varphi\varphi SS} c_\alpha^2,
\end{eqnarray}
where $c_\alpha=\cos{\alpha}$ and $s_\alpha=\sin{\alpha}$. In the limit $\alpha\rightarrow0$, 
using Eqns.~\ref{eqn:lambdaphi} and \ref{eqn:lambdaH},
these vertices become
\begin{eqnarray}
g_{h_1h_1h_1} & = & 3i\frac{m_{h_1}^2}{v} \\
g_{h_2h_2h_2} & = & 3i\frac{m_{h_2}^2}{w}\\
g_{h_2SS} & = & 2i\lambda_{\varphi\varphi SS}w\\
g_{h_2h_2SS} & = & 2i\lambda_{\varphi\varphi SS},
\end{eqnarray}
with all other couplings going to 0, effectively decoupling $h_1$ from the other scalars.

Under this approximation, the amplitudes of $SS\rightarrow h_1h_1$ and $SS\rightarrow h_1h_2$ vanish, 
and the amplitude of $SS\rightarrow h_2h_2$ is given by
\begin{eqnarray}
\mathcal{M}_{SS\rightarrow h_2h_2} & = & g_{h_2h_2SS}-\frac{ig_{h_2h_2h_2}g_{h_2SS}}{s-m_{h_2}^2}\nonumber\\
&& -\frac{ig_{h_2SS}^2}{t-m_S^2}-\frac{ig_{h_2SS}^2}{u-m_S^2}\\
& \approx & 4i \lambda_{\varphi\varphi SS} - 8i\frac{w^2\lambda_{\varphi\varphi SS}^2}{m_S^2},
\end{eqnarray}
where we used a theshold approximation to set the Mandelstam variables to 
$s=\left(2m_{h_2}\right)^2$ and $t=u=0$. The resulting amplitude is zero at both $\lambda_{\varphi\varphi SS}=0$ 
and $\lambda_{\varphi\varphi SS}=m_S^2/2w^2$.

\end{appendices}


\begin{thebibliography}{99} 

%1
%\cite{Bergstrom:2000pn}
\bibitem{Bergstrom:2000pn}
L.~Bergstr\"om,
Nonbaryonic dark matter: Observational evidence and detection methods,
Rept. Prog. Phys. \textbf{63}, 793 (2000)
doi:10.1088/0034-4885/63/5/2r3
[arXiv:hep-ph/0002126 [hep-ph]].

%2
%\cite{Bertone:2004pz}
\bibitem{Bertone:2004pz}
G.~Bertone, D.~Hooper and J.~Silk,
Particle dark matter: Evidence, candidates and constraints,
Phys. Rept. \textbf{405}, 279-390 (2005)
doi:10.1016/j.physrep.2004.08.031
[arXiv:hep-ph/0404175 [hep-ph]].

%3
%\cite{Bergstrom:2012fi}
\bibitem{Bergstrom:2012fi}
L.~Bergstrom,
Dark Matter Evidence, Particle Physics Candidates and Detection Methods,
Annalen Phys. \textbf{524}, 479-496 (2012)
doi:10.1002/andp.201200116
[arXiv:1205.4882 [astro-ph.HE]].

%4
%\cite{Silveira:1985rk}
\bibitem{Silveira:1985rk}
V.~Silveira and A.~Zee,
SCALAR PHANTOMS,
Phys. Lett. B \textbf{161}, 136-140 (1985)
doi:10.1016/0370-2693(85)90624-0

%5
%\cite{McDonald:1993ex}
\bibitem{McDonald:1993ex}
J.~McDonald,
Gauge singlet scalars as cold dark matter,
Phys. Rev. D \textbf{50}, 3637-3649 (1994)
doi:10.1103/PhysRevD.50.3637
[arXiv:hep-ph/0702143 [hep-ph]].

%6
%\cite{Patt:2006fw}
\bibitem{Patt:2006fw}
B.~Patt and F.~Wilczek,
Higgs-field portal into hidden sectors,
[arXiv:hep-ph/0605188 [hep-ph]].

%7
%\cite{Baek:2011aa}
\bibitem{Baek:2011aa}
S.~Baek, P.~Ko and W.~I.~Park,
Search for the Higgs portal to a singlet fermionic dark matter at the LHC,
JHEP \textbf{02}, 047 (2012)
doi:10.1007/JHEP02(2012)047
[arXiv:1112.1847 [hep-ph]].

%8
%\cite{Djouadi:2011aa}
\bibitem{Djouadi:2011aa}
A.~Djouadi, O.~Lebedev, Y.~Mambrini and J.~Quevillon,
Implications of LHC searches for Higgs--portal dark matter,
Phys. Lett. B \textbf{709}, 65-69 (2012)
doi:10.1016/j.physletb.2012.01.062
[arXiv:1112.3299 [hep-ph]].

%9
%\cite{LopezHonorez:2012kv}
\bibitem{LopezHonorez:2012kv}
L.~Lopez-Honorez, T.~Schwetz and J.~Zupan,
Higgs portal, fermionic dark matter, and a Standard Model like Higgs at 125 GeV,
Phys. Lett. B \textbf{716}, 179-185 (2012)
doi:10.1016/j.physletb.2012.07.017
[arXiv:1203.2064 [hep-ph]].

%10
%\cite{Baek:2012se}
\bibitem{Baek:2012se}
S.~Baek, P.~Ko, W.~I.~Park and E.~Senaha,
Higgs Portal Vector Dark Matter : Revisited,
JHEP \textbf{05}, 036 (2013)
doi:10.1007/JHEP05(2013)036
[arXiv:1212.2131 [hep-ph]].

%11
%\cite{Walker:2013hka}
\bibitem{Walker:2013hka}
D.~G.~E.~Walker,
Unitarity Constraints on Higgs Portals,
[arXiv:1310.1083 [hep-ph]].

%12
%\cite{Esch:2014jpa}
\bibitem{Esch:2014jpa}
S.~Esch, M.~Klasen and C.~E.~Yaguna,
A minimal model for two-component dark matter,
JHEP \textbf{09}, 108 (2014)
doi:10.1007/JHEP09(2014)108
[arXiv:1406.0617 [hep-ph]].

%13
%\cite{Buchmueller:2014yoa}
\bibitem{Buchmueller:2014yoa}
O.~Buchmueller, M.~J.~Dolan, S.~A.~Malik and C.~McCabe,
Characterising dark matter searches at colliders and direct detection experiments: Vector mediators,
JHEP \textbf{01}, 037 (2015)
doi:10.1007/JHEP01(2015)037
[arXiv:1407.8257 [hep-ph]].

%14
%\cite{Cheung:2015dta}
\bibitem{Cheung:2015dta}
K.~Cheung, P.~Ko, J.~S.~Lee and P.~Y.~Tseng,
Bounds on Higgs-Portal models from the LHC Higgs data,
JHEP \textbf{10}, 057 (2015)
doi:10.1007/JHEP10(2015)057
[arXiv:1507.06158 [hep-ph]].

%15
%\cite{Beniwal:2015sdl}
\bibitem{Beniwal:2015sdl}
A.~Beniwal, F.~Rajec, C.~Savage, P.~Scott, C.~Weniger, M.~White and A.~G.~Williams,
Combined analysis of effective Higgs portal dark matter models,
Phys. Rev. D \textbf{93}, no.11, 115016 (2016)
doi:10.1103/PhysRevD.93.115016
[arXiv:1512.06458 [hep-ph]].

%16
%\cite{Han:2016gyy}
\bibitem{Han:2016gyy}
H.~Han, J.~M.~Yang, Y.~Zhang and S.~Zheng,
Collider Signatures of Higgs-portal Scalar Dark Matter,
Phys. Lett. B \textbf{756}, 109-112 (2016)
doi:10.1016/j.physletb.2016.03.010
[arXiv:1601.06232 [hep-ph]].

%17
%\cite{Arcadi:2016qoz}
\bibitem{Arcadi:2016qoz}
G.~Arcadi, C.~Gross, O.~Lebedev, S.~Pokorski and T.~Toma,
Evading Direct Dark Matter Detection in Higgs Portal Models,
Phys. Lett. B \textbf{769}, 129-133 (2017)
doi:10.1016/j.physletb.2017.03.044
[arXiv:1611.09675 [hep-ph]].

%18
%\cite{Arcadi:2017kky}
\bibitem{Arcadi:2017kky}
G.~Arcadi, M.~Dutra, P.~Ghosh, M.~Lindner, Y.~Mambrini, M.~Pierre, S.~Profumo and F.~S.~Queiroz,
The waning of the WIMP? A review of models, searches, and constraints,
Eur. Phys. J. C \textbf{78}, no.3, 203 (2018)
doi:10.1140/epjc/s10052-018-5662-y
[arXiv:1703.07364 [hep-ph]].

%19
%\cite{Bhattacharya:2017fid}
\bibitem{Bhattacharya:2017fid}
S.~Bhattacharya, P.~Ghosh, T.~N.~Maity and T.~S.~Ray,
Mitigating Direct Detection Bounds in Non-minimal Higgs Portal Scalar Dark Matter Models,
JHEP \textbf{10}, 088 (2017)
doi:10.1007/JHEP10(2017)088
[arXiv:1706.04699 [hep-ph]].

%20
%\cite{Gross:2017dan}
\bibitem{Gross:2017dan}
C.~Gross, O.~Lebedev and T.~Toma,
Cancellation Mechanism for Dark-Matter-Nucleon Interaction,
Phys. Rev. Lett. \textbf{119}, no.19, 191801 (2017)
doi:10.1103/PhysRevLett.119.191801
[arXiv:1708.02253 [hep-ph]].

%21
%\cite{Azevedo:2018oxv}
\bibitem{Azevedo:2018oxv}
D.~Azevedo, M.~Duch, B.~Grzadkowski, D.~Huang, M.~Iglicki and R.~Santos,
Testing scalar versus vector dark matter,
Phys. Rev. D \textbf{99}, no.1, 015017 (2019)
doi:10.1103/PhysRevD.99.015017
[arXiv:1808.01598 [hep-ph]].

%22
%\cite{Arcadi:2019lka}
\bibitem{Arcadi:2019lka}
G.~Arcadi, A.~Djouadi and M.~Raidal,
Dark Matter through the Higgs portal,'
Phys. Rept. \textbf{842}, 1-180 (2020)
doi:10.1016/j.physrep.2019.11.003
[arXiv:1903.03616 [hep-ph]].

%23
%\cite{Cabrera:2019gaq}
\bibitem{Cabrera:2019gaq}
M.~Cabrera, J.~Casas, A.~Delgado and S.~Robles,
Generalized Blind Spots for Dark Matter Direct Detection in the 2HDM,
JHEP \textbf{02}, 166 (2020)
doi:10.1007/JHEP02(2020)166
[arXiv:1912.01758 [hep-ph]].

%\cite{Ghorbani:2014gka}
\bibitem{Ghorbani:2014gka}
K.~Ghorbani and H.~Ghorbani,
Scalar split WIMPs in future direct detection experiments,
Phys. Rev. D \textbf{93}, no.5, 055012 (2016)
doi:10.1103/PhysRevD.93.055012
[arXiv:1501.00206 [hep-ph]].

%\cite{Ghosh:2017fmr}
\bibitem{Ghosh:2017fmr}
P.~Ghosh, A.~K.~Saha and A.~Sil,
Study of Electroweak Vacuum Stability from Extended Higgs Portal of Dark Matter and Neutrinos,
Phys. Rev. D \textbf{97}, no.7, 075034 (2018)
doi:10.1103/PhysRevD.97.075034
[arXiv:1706.04931 [hep-ph]].

%\cite{Alanne:2020jwx}
\bibitem{Alanne:2020jwx}
T.~Alanne, N.~Benincasa, M.~Heikinheimo, K.~Kannike, V.~Keus, N.~Koivunen and K.~Tuominen,
Pseudo-Goldstone dark matter: gravitational waves and direct-detection blind spots,
JHEP \textbf{10}, 080 (2020)
doi:10.1007/JHEP10(2020)080
[arXiv:2008.09605 [hep-ph]].

%24
%\cite{Buckley:2011mm}
\bibitem{Buckley:2011mm}
M.~R.~Buckley, D.~Hooper and J.~L.~Rosner,
A Leptophobic Z' And Dark Matter From Grand Unification,
Phys. Lett. B \textbf{703}, 343-347 (2011)
doi:10.1016/j.physletb.2011.08.014
[arXiv:1106.3583 [hep-ph]].

%25
%\cite{Frandsen:2011cg}
\bibitem{Frandsen:2011cg}
M.~T.~Frandsen, F.~Kahlhoefer, S.~Sarkar and K.~Schmidt-Hoberg,
Direct detection of dark matter in models with a light Z',
JHEP \textbf{09}, 128 (2011)
doi:10.1007/JHEP09(2011)128
[arXiv:1107.2118 [hep-ph]].

%26
%\cite{Lebedev:2011iq}
\bibitem{Lebedev:2011iq}
O.~Lebedev, H.~M.~Lee and Y.~Mambrini,
Vector Higgs-portal dark matter and the invisible Higgs,'
Phys. Lett. B \textbf{707}, 570-576 (2012)
doi:10.1016/j.physletb.2012.01.029
[arXiv:1111.4482 [hep-ph]].

%27
%\cite{Alves:2013tqa}
\bibitem{Alves:2013tqa}
A.~Alves, S.~Profumo and F.~S.~Queiroz,
The dark $Z^{'}$ portal: direct, indirect and collider searches,'
JHEP \textbf{04}, 063 (2014)
doi:10.1007/JHEP04(2014)063
[arXiv:1312.5281 [hep-ph]].

%28
%\cite{Arcadi:2013qia}
\bibitem{Arcadi:2013qia}
G.~Arcadi, Y.~Mambrini, M.~H.~G.~Tytgat and B.~Zaldivar,
Invisible $Z^\prime$ and dark matter: LHC vs LUX constraints,
JHEP \textbf{03}, 134 (2014)
doi:10.1007/JHEP03(2014)134
[arXiv:1401.0221 [hep-ph]].

%29
%\cite{Lebedev:2014bba}
\bibitem{Lebedev:2014bba}
O.~Lebedev and Y.~Mambrini,
Axial dark matter: The case for an invisible $Z'$,
Phys. Lett. B \textbf{734}, 350-353 (2014)
doi:10.1016/j.physletb.2014.05.025
[arXiv:1403.4837 [hep-ph]].

%30
%\cite{Arcadi:2014lta}
\bibitem{Arcadi:2014lta}
G.~Arcadi, Y.~Mambrini and F.~Richard,
Z-portal dark matter,'
JCAP \textbf{03}, 018 (2015)
doi:10.1088/1475-7516/2015/03/018
[arXiv:1411.2985 [hep-ph]].

%31
%\cite{Hooper:2014fda}
\bibitem{Hooper:2014fda}
D.~Hooper,
$Z^\prime$ Mediated Dark Matter Models for the Galactic Center Gamma-Ray Excess,
Phys. Rev. D \textbf{91}, 035025 (2015)
doi:10.1103/PhysRevD.91.035025
[arXiv:1411.4079 [hep-ph]].

%32
%\cite{Alves:2015pea}
\bibitem{Alves:2015pea}
A.~Alves, A.~Berlin, S.~Profumo and F.~S.~Queiroz,
Dark Matter Complementarity and the Z$^\prime$ Portal,
Phys. Rev. D \textbf{92}, no.8, 083004 (2015)
doi:10.1103/PhysRevD.92.083004
[arXiv:1501.03490 [hep-ph]].

%33
%\cite{Ghorbani:2015baa}
\bibitem{Ghorbani:2015baa}
K.~Ghorbani and H.~Ghorbani,
Two-portal Dark Matter,
Phys. Rev. D \textbf{91}, no.12, 123541 (2015)
doi:10.1103/PhysRevD.91.123541
[arXiv:1504.03610 [hep-ph]].

%34
%\cite{Jacques:2016dqz}
\bibitem{Jacques:2016dqz}
T.~Jacques, A.~Katz, E.~Morgante, D.~Racco, M.~Rameez and A.~Riotto,
Complementarity of DM searches in a consistent simplified model: the case of $Z'$',
JHEP \textbf{10}, 071 (2016)
[erratum: JHEP \textbf{01}, 127 (2019)]
doi:10.1007/JHEP10(2016)071
[arXiv:1605.06513 [hep-ph]].

%35
%\cite{Duerr:2016tmh}
\bibitem{Duerr:2016tmh}
M.~Duerr, F.~Kahlhoefer, K.~Schmidt-Hoberg, T.~Schwetz and S.~Vogl,
How to save the WIMP: global analysis of a dark matter model with two s-channel mediators,
JHEP \textbf{09}, 042 (2016)
doi:10.1007/JHEP09(2016)042
[arXiv:1606.07609 [hep-ph]].

%36
%\cite{Ismail:2016tod}
\bibitem{Ismail:2016tod}
A.~Ismail, W.~Y.~Keung, K.~H.~Tsao and J.~Unwin,
Axial vector $Z'$ and anomaly cancellation,
Nucl. Phys. B \textbf{918}, 220-244 (2017)
doi:10.1016/j.nuclphysb.2017.03.001
[arXiv:1609.02188 [hep-ph]].

%37
%\cite{Escudero:2016gzx}
\bibitem{Escudero:2016gzx}
M.~Escudero, A.~Berlin, D.~Hooper and M.~X.~Lin,
Toward (Finally!) Ruling Out Z and Higgs Mediated Dark Matter Models,
JCAP \textbf{12}, 029 (2016)
doi:10.1088/1475-7516/2016/12/029
[arXiv:1609.09079 [hep-ph]].

%38
%\cite{Kearney:2016rng}
\bibitem{Kearney:2016rng}
J.~Kearney, N.~Orlofsky and A.~Pierce,
$Z$ boson mediated dark matter beyond the effective theory,
Phys. Rev. D \textbf{95}, no.3, 035020 (2017)
doi:10.1103/PhysRevD.95.035020
[arXiv:1611.05048 [hep-ph]].

%39
%\cite{Alves:2016cqf}
\bibitem{Alves:2016cqf}
A.~Alves, G.~Arcadi, Y.~Mambrini, S.~Profumo and F.~S.~Queiroz,
Augury of darkness: the low-mass dark $Z'$ portal,
JHEP \textbf{04}, 164 (2017)
doi:10.1007/JHEP04(2017)164
[arXiv:1612.07282 [hep-ph]].

%40
%\cite{Dutra:2018gmv}
\bibitem{Dutra:2018gmv}
M.~Dutra, M.~Lindner, S.~Profumo, F.~S.~Queiroz, W.~Rodejohann and C.~Siqueira,
MeV Dark Matter Complementarity and the Dark Photon Portal,
JCAP \textbf{03}, 037 (2018)
doi:10.1088/1475-7516/2018/03/037
[arXiv:1801.05447 [hep-ph]].

%41
%\cite{Okada:2018ktp}
\bibitem{Okada:2018ktp}
S.~Okada,
$Z'$ Portal Dark Matter in the Minimal $B-L$ Model,
Adv. High Energy Phys. \textbf{2018}, 5340935 (2018)
doi:10.1155/2018/5340935
[arXiv:1803.06793 [hep-ph]].

%42
%\cite{Blanco:2019hah}
\bibitem{Blanco:2019hah}
C.~Blanco, M.~Escudero, D.~Hooper and S.~J.~Witte,
Z' mediated WIMPs: dead, dying, or soon to be detected?,
JCAP \textbf{11}, 024 (2019)
doi:10.1088/1475-7516/2019/11/024
[arXiv:1907.05893 [hep-ph]].

%43
%\cite{Arcadi:2020jqf}
\bibitem{Arcadi:2020jqf}
G.~Arcadi, A.~Djouadi and M.~Kado,
The Higgs-portal for vector Dark Matter and the Effective Field Theory approach: a reappraisal,
Phys. Lett. B \textbf{805}, 135427 (2020)
doi:10.1016/j.physletb.2020.135427
[arXiv:2001.10750 [hep-ph]].

%44
%\cite{Okada:2020cue}
\bibitem{Okada:2020cue}
N.~Okada, S.~Okada and Q.~Shafi,
Light $Z'$ and dark matter from U(1)$_X$ gauge symmetry,
Phys. Lett. B \textbf{810}, 135845 (2020)
doi:10.1016/j.physletb.2020.135845
[arXiv:2003.02667 [hep-ph]].

%45
%\cite{Falkowski:2009yz}
\bibitem{Falkowski:2009yz}
A.~Falkowski, J.~Juknevich and J.~Shelton,
Dark Matter Through the Neutrino Portal,
[arXiv:0908.1790 [hep-ph]].

%46
%\cite{Cherry:2014xra}
\bibitem{Cherry:2014xra}
J.~F.~Cherry, A.~Friedland and I.~M.~Shoemaker,
Neutrino Portal Dark Matter: From Dwarf Galaxies to IceCube,
[arXiv:1411.1071 [hep-ph]].

%47
%\cite{Batell:2017cmf}
\bibitem{Batell:2017cmf}
B.~Batell, T.~Han, D.~McKeen and B.~Shams Es Haghi,
Thermal Dark Matter Through the Dirac Neutrino Portal,
Phys. Rev. D \textbf{97}, no.7, 075016 (2018)
doi:10.1103/PhysRevD.97.075016
[arXiv:1709.07001 [hep-ph]].

%48
%\cite{Cosme:2020mck}
\bibitem{Cosme:2020mck}
C.~Cosme, M.~Dutra, T.~Ma, Y.~Wu and L.~Yang,
Neutrino Portal to FIMP Dark Matter with an Early Matter Era,
JHEP \textbf{21}, 026 (2020)
doi:10.1007/JHEP03(2021)026
[arXiv:2003.01723 [hep-ph]].

%49
%\cite{Cheung:2012qy}
\bibitem{Cheung:2012qy}
C.~Cheung, L.~J.~Hall, D.~Pinner and J.~T.~Ruderman,
Prospects and Blind Spots for Neutralino Dark Matter,
JHEP \textbf{05}, 100 (2013)
doi:10.1007/JHEP05(2013)100
[arXiv:1211.4873 [hep-ph]].

%50
%\cite{Cheung:2013dua}
\bibitem{Cheung:2013dua}
C.~Cheung and D.~Sanford,
Simplified Models of Mixed Dark Matter,
JCAP \textbf{02}, 011 (2014)
doi:10.1088/1475-7516/2014/02/011
[arXiv:1311.5896 [hep-ph]].

%51
%\cite{Huang:2014xua}
\bibitem{Huang:2014xua}
P.~Huang and C.~E.~M.~Wagner,
Blind Spots for neutralino Dark Matter in the MSSM with an intermediate$ m_A$,
Phys. Rev. D \textbf{90}, no.1, 015018 (2014)
doi:10.1103/PhysRevD.90.015018
[arXiv:1404.0392 [hep-ph]].

%52
%\cite{Berlin:2015wwa}
\bibitem{Berlin:2015wwa}
A.~Berlin, S.~Gori, T.~Lin and L.~T.~Wang,
Pseudoscalar Portal Dark Matter,
Phys. Rev. D \textbf{92}, 015005 (2015)
doi:10.1103/PhysRevD.92.015005
[arXiv:1502.06000 [hep-ph]].

%53
%\cite{Casas:2017jjg}
\bibitem{Casas:2017jjg}
J.~A.~Casas, D.~G.~Cerde\~no, J.~M.~Moreno and J.~Quilis,
Reopening the Higgs portal for single scalar dark matter,
JHEP \textbf{05}, 036 (2017)
doi:10.1007/JHEP05(2017)036
[arXiv:1701.08134 [hep-ph]].
 
 %54
%\cite{Huitu:2018gbc}
\bibitem{Huitu:2018gbc}
K.~Huitu, N.~Koivunen, O.~Lebedev, S.~Mondal and T.~Toma,
Probing pseudo-Goldstone dark matter at the LHC,'
Phys. Rev. D \textbf{100}, no.1, 015009 (2019)
doi:10.1103/PhysRevD.100.015009
[arXiv:1812.05952 [hep-ph]].

%55
%\cite{Barger:2007im}
\bibitem{Barger:2007im}
V.~Barger, P.~Langacker, M.~McCaskey, M.~J.~Ramsey-Musolf and G.~Shaughnessy,
LHC Phenomenology of an Extended Standard Model with a Real Scalar Singlet,
Phys. Rev. D \textbf{77}, 035005 (2008)
doi:10.1103/PhysRevD.77.035005
[arXiv:0706.4311 [hep-ph]].

%56
%\cite{Falkowski:2015iwa}
\bibitem{Falkowski:2015iwa}
A.~Falkowski, C.~Gross and O.~Lebedev,
A second Higgs from the Higgs portal,
JHEP \textbf{05}, 057 (2015)
doi:10.1007/JHEP05(2015)057
[arXiv:1502.01361 [hep-ph]].

%57
%\cite{Ko:2016ybp}
\bibitem{Ko:2016ybp}
P.~Ko and J.~Li,
Interference effects of two scalar boson propagators on the LHC search for the singlet fermion DM,
Phys. Lett. B \textbf{765}, 53-61 (2017)
doi:10.1016/j.physletb.2016.11.056
[arXiv:1610.03997 [hep-ph]].

%58
%\cite{Arcadi:2016kmk}
\bibitem{Arcadi:2016kmk}
G.~Arcadi, C.~Gross, O.~Lebedev, Y.~Mambrini, S.~Pokorski and T.~Toma,
Multicomponent Dark Matter from Gauge Symmetry,
JHEP \textbf{12}, 081 (2016)
doi:10.1007/JHEP12(2016)081
[arXiv:1611.00365 [hep-ph]].

%59
%\cite{Bell:2016ekl}
\bibitem{Bell:2016ekl}
N.~F.~Bell, G.~Busoni and I.~W.~Sanderson,
Self-consistent Dark Matter Simplified Models with an s-channel scalar mediator,
JCAP \textbf{03}, 015 (2017)
doi:10.1088/1475-7516/2017/03/015
[arXiv:1612.03475 [hep-ph]].

%60
%\cite{Bell:2017rgi}
\bibitem{Bell:2017rgi}
N.~F.~Bell, G.~Busoni and I.~W.~Sanderson,
Two Higgs Doublet Dark Matter Portal,
JCAP \textbf{01}, 015 (2018)
doi:10.1088/1475-7516/2018/01/015
[arXiv:1710.10764 [hep-ph]].

%61
%\cite{Arcadi:2018pfo}
\bibitem{Arcadi:2018pfo}
G.~Arcadi,
2HDM portal for Singlet-Doublet Dark Matter,
Eur. Phys. J. C \textbf{78}, no.10, 864 (2018)
doi:10.1140/epjc/s10052-018-6327-6
[arXiv:1804.04930 [hep-ph]].

%62
%\cite{Buckley:2014fba}
\bibitem{Buckley:2014fba}
M.~R.~Buckley, D.~Feld and D.~Goncalves,
Scalar Simplified Models for Dark Matter,'
Phys. Rev. D \textbf{91}, 015017 (2015)
doi:10.1103/PhysRevD.91.015017
[arXiv:1410.6497 [hep-ph]].

%63
%\cite{No:2015xqa}
\bibitem{No:2015xqa}
J.~M.~No,
Looking through the pseudoscalar portal into dark matter: Novel mono-Higgs and mono-Z signatures at the LHC,
Phys. Rev. D \textbf{93}, no.3, 031701 (2016)
doi:10.1103/PhysRevD.93.031701
[arXiv:1509.01110 [hep-ph]].

%64
%\cite{Goncalves:2016iyg}
\bibitem{Goncalves:2016iyg}
D.~Goncalves, P.~A.~N.~Machado and J.~M.~No,
Simplified Models for Dark Matter Face their Consistent Completions,
Phys. Rev. D \textbf{95}, no.5, 055027 (2017)
doi:10.1103/PhysRevD.95.055027
[arXiv:1611.04593 [hep-ph]].

%65
%\cite{Tunney:2017yfp}
\bibitem{Tunney:2017yfp}
P.~Tunney, J.~M.~No and M.~Fairbairn,
Probing the pseudoscalar portal to dark matter via $\bar bbZ(\to\ell\ell)+ \not{E}_T$ : From the LHC to the Galactic Center excess,
Phys. Rev. D \textbf{96}, no.9, 095020 (2017)
doi:10.1103/PhysRevD.96.095020
[arXiv:1705.09670 [hep-ph]].

%66
%\cite{Arcadi:2017wqi}
\bibitem{Arcadi:2017wqi}
G.~Arcadi, M.~Lindner, F.~S.~Queiroz, W.~Rodejohann and S.~Vogl,
Pseudoscalar Mediators: A WIMP model at the Neutrino Floor,
JCAP \textbf{03}, 042 (2018)
doi:10.1088/1475-7516/2018/03/042
[arXiv:1711.02110 [hep-ph]].

%67
%\cite{Ghosh:2020fdc}
\bibitem{Ghosh:2020fdc}
S.~Ghosh, A.~Dutta Banik, E.~J.~Chun and D.~Majumdar,
Pseudoscalar-portal Dark Matter in the Light of AMS-02 positron excess,'
[arXiv:2003.07675 [hep-ph]].

%68
%\cite{Butterworth:2020vnb}
\bibitem{Butterworth:2020vnb}
J.~M.~Butterworth, M.~Habedank, P.~Pani and A.~Vaitkus,
A study of collider signatures for two Higgs doublet models with a Pseudoscalar mediator to Dark Matter,'
[arXiv:2009.02220 [hep-ph]].

%69
%\cite{Abe:2020pzp}
\bibitem{Abe:2020pzp}
T.~Abe, M.~Fujiwara, J.~Hisano and Y.~Shoji,
Future detectability of a pseudoscalar mediator dark matter model,'
J. Phys. Conf. Ser. \textbf{1468}, no.1, 012012 (2020)
doi:10.1088/1742-6596/1468/1/012012

%70
%\cite{Okada:2020zxo}
\bibitem{Okada:2020zxo}
N.~Okada, D.~Raut and Q.~Shafi,
Pseudo-Goldstone Dark Matter in gauged $B-L$ extended Standard Model,'
[arXiv:2001.05910 [hep-ph]].

%%71
%%\cite{Alanne:2020jwx}
%\bibitem{Alanne:2020jwx}
%T.~Alanne, N.~Benincasa, M.~Heikinheimo, K.~Kannike, V.~Keus, N.~Koivunen and K.~Tuominen,
%Pseudo-Goldstone dark matter: gravitational waves and direct-detection blind spots,'
%JHEP \textbf{10}, 080 (2020)
%doi:10.1007/JHEP10(2020)080
%[arXiv:2008.09605 [hep-ph]].

%72
%\cite{Zhang:2021alu}
\bibitem{Zhang:2021alu}
Z.~Zhang, C.~Cai, X.~M.~Jiang, Y.~L.~Tang, Z.~H.~Yu and H.~H.~Zhang,
Phase transition gravitational waves from pseudo-Nambu-Goldstone dark matter and two Higgs doublets,'
[arXiv:2102.01588 [hep-ph]].

%73
%\cite{Bauer:2017ota}
\bibitem{Bauer:2017ota}
M.~Bauer, U.~Haisch and F.~Kahlhoefer,
Simplified dark matter models with two Higgs doublets: I. Pseudoscalar mediators,'
JHEP \textbf{05}, 138 (2017)
doi:10.1007/JHEP05(2017)138
[arXiv:1701.07427 [hep-ph]].

%74
%\cite{Jiang:2019soj}
\bibitem{Jiang:2019soj}
X.~M.~Jiang, C.~Cai, Z.~H.~Yu, Y.~P.~Zeng and H.~H.~Zhang,
Pseudo-Nambu-Goldstone dark matter and two-Higgs-doublet models,'
Phys. Rev. D \textbf{100}, no.7, 075011 (2019)
doi:10.1103/PhysRevD.100.075011
[arXiv:1907.09684 [hep-ph]].

%75
%\cite{Arcadi:2020gge}
\bibitem{Arcadi:2020gge}
G.~Arcadi, G.~Busoni, T.~Hugle and V.~T.~Tenorth,
Comparing 2HDM $+$ Scalar and Pseudoscalar Simplified Models at LHC,'
JHEP \textbf{06}, 098 (2020)
doi:10.1007/JHEP06(2020)098
[arXiv:2001.10540 [hep-ph]].

%76
%\cite{Arhrib:2011uy}
\bibitem{Arhrib:2011uy}
A.~Arhrib, R.~Benbrik, M.~Chabab, G.~Moultaka, M.~C.~Peyranere, L.~Rahili and J.~Ramadan,
The Higgs Potential in the Type II Seesaw Model,
Phys. Rev. D \textbf{84}, 095005 (2011)
doi:10.1103/PhysRevD.84.095005
[arXiv:1105.1925 [hep-ph]].

%\cite{Ellis:2005mb}
\bibitem{Ellis:2005mb}
J.~R.~Ellis, K.~A.~Olive, Y.~Santoso and V.~C.~Spanos,
Update on the direct detection of supersymmetric dark matter,
Phys. Rev. D \textbf{71}, 095007 (2005)
doi:10.1103/PhysRevD.71.095007
[arXiv:hep-ph/0502001 [hep-ph]].

%\cite{Carena:2006nv}
\bibitem{Carena:2006nv}
M.~Carena, D.~Hooper and A.~Vallinotto,
The Interplay Between Collider Searches For Supersymmetric Higgs Bosons and Direct Dark Matter Experiments,
Phys. Rev. D \textbf{75}, 055010 (2007)
doi:10.1103/PhysRevD.75.055010
[arXiv:hep-ph/0611065 [hep-ph]].

%\cite{Hooper:2006wv}
\bibitem{Hooper:2006wv}
D.~Hooper and A.~M.~Taylor,
Determining Supersymmetric Parameters With Dark Matter Experiments,
JCAP \textbf{03}, 017 (2007)
doi:10.1088/1475-7516/2007/03/017
[arXiv:hep-ph/0607086 [hep-ph]].

%\cite{Cao:2010ph}
\bibitem{Cao:2010ph}
J.~Cao, K.~i.~Hikasa, W.~Wang, J.~M.~Yang and L.~X.~Yu,
Constraints of dark matter direct detection experiments on the MSSM and implications on LHC Higgs search,
Phys. Rev. D \textbf{82}, 051701 (2010)
doi:10.1103/PhysRevD.82.051701
[arXiv:1006.4811 [hep-ph]].



%77
%\cite{Campbell:2016zbp}
\bibitem{Campbell:2016zbp}
R.~Campbell, S.~Godfrey, H.~E.~Logan and A.~Poulin,
Real singlet scalar dark matter extension of the Georgi-Machacek model,
Phys. Rev. D \textbf{95}, no.1, 016005 (2017)
doi:10.1103/PhysRevD.95.016005
[arXiv:1610.08097 [hep-ph]].

%78
%\cite{Belanger:2018ccd}
\bibitem{Belanger:2018ccd}
G.~B\'elanger, F.~Boudjema, A.~Goudelis, A.~Pukhov and B.~Zaldivar,
micrOMEGAs5.0 : Freeze-in,
Comput. Phys. Commun. \textbf{231}, 173-186 (2018)
doi:10.1016/j.cpc.2018.04.027
[arXiv:1801.03509 [hep-ph]].

%79
%\cite{Tanabashi:2018oca}
\bibitem{Tanabashi:2018oca}
P.A. Zyla \textit{et al.} [Particle Data Group],
Review of Particle Physics,
 Prog. Theor. Exp. Phys. 2020, 083C01 (2020).
 
%80
%\cite{Aaboud:2019rtt}
\bibitem{Aaboud:2019rtt}
M.~Aaboud \textit{et al.} [ATLAS],
Combination of searches for invisible Higgs boson decays with the ATLAS experiment,
Phys. Rev. Lett. \textbf{122}, no.23, 231801 (2019)
doi:10.1103/PhysRevLett.122.231801
[arXiv:1904.05105 [hep-ex]].

%81
%\cite{Sirunyan:2018owy}
\bibitem{Sirunyan:2018owy}
A.~M.~Sirunyan \textit{et al.} [CMS],
Search for invisible decays of a Higgs boson produced through vector boson fusion in proton-proton collisions at $\sqrt{s} =$ 13 TeV,
Phys. Lett. B \textbf{793}, 520-551 (2019)
doi:10.1016/j.physletb.2019.04.025
[arXiv:1809.05937 [hep-ex]].

%82
%\cite{Djouadi:1997yw}
\bibitem{Djouadi:1997yw}
A.~Djouadi, J.~Kalinowski and M.~Spira,
HDECAY: A Program for Higgs boson decays in the standard model and its supersymmetric extension,
Comput. Phys. Commun. \textbf{108}, 56-74 (1998)
doi:10.1016/S0010-4655(97)00123-9
[arXiv:hep-ph/9704448 [hep-ph]].

%83
%\cite{Bechtle:2013xfa}
\bibitem{Bechtle:2013xfa}
P.~Bechtle, S.~Heinemeyer, O.~Stål, T.~Stefaniak and G.~Weiglein,
$HiggsSignals$: Confronting arbitrary Higgs sectors with measurements at the Tevatron and the LHC,
Eur. Phys. J. C \textbf{74}, no.2, 2711 (2014)
doi:10.1140/epjc/s10052-013-2711-4
[arXiv:1305.1933 [hep-ph]].

%84
%\cite{Robertson:2016xjh}
\bibitem{Robertson:2016xjh}
A.~Robertson, R.~Massey and V.~Eke,
What does the Bullet Cluster tell us about self-interacting dark matter?,
Mon. Not. Roy. Astron. Soc. \textbf{465}, no.1, 569-587 (2017)
doi:10.1093/mnras/stw2670
[arXiv:1605.04307 [astro-ph.CO]].

%85
%\cite{Harvey:2015hha}
\bibitem{Harvey:2015hha}
D.~Harvey, R.~Massey, T.~Kitching, A.~Taylor and E.~Tittley,
The non-gravitational interactions of dark matter in colliding galaxy clusters,
Science \textbf{347}, 1462-1465 (2015)
doi:10.1126/science.1261381
[arXiv:1503.07675 [astro-ph.CO]].

%86
%\cite{Hoof:2018hyn}
\bibitem{Hoof:2018hyn}
S.~Hoof, A.~Geringer-Sameth and R.~Trotta,
A Global Analysis of Dark Matter Signals from 27 Dwarf Spheroidal Galaxies using 11 Years of Fermi-LAT Observations,
JCAP \textbf{02}, 012 (2020)
doi:10.1088/1475-7516/2020/02/012
[arXiv:1812.06986 [astro-ph.CO]].

%87
%\cite{Atwood:2009ez}
\bibitem{Atwood:2009ez}
W.~B.~Atwood \textit{et al.} [Fermi-LAT],
The Large Area Telescope on the Fermi Gamma-ray Space Telescope Mission,
Astrophys. J. \textbf{697}, 1071-1102 (2009)
doi:10.1088/0004-637X/697/2/1071
[arXiv:0902.1089 [astro-ph.IM]].

%88
%\cite{Kolb:1990vq}
\bibitem{Kolb:1990vq}
E.~W.~Kolb and M.~S.~Turner,
The Early Universe,
Front. Phys. \textbf{69}, 1-547 (1990)

%89
%\cite{Aprile:2017iyp}
\bibitem{Aprile:2017iyp}
E.~Aprile \textit{et al.} [XENON],
First Dark Matter Search Results from the XENON1T Experiment,
Phys. Rev. Lett. \textbf{119}, no.18, 181301 (2017)
doi:10.1103/PhysRevLett.119.181301
[arXiv:1705.06655 [astro-ph.CO]].

%90
%\cite{Cline:2013gha}
\bibitem{Cline:2013gha}
J.~M.~Cline, K.~Kainulainen, P.~Scott and C.~Weniger,
Update on scalar singlet dark matter,
Phys. Rev. D \textbf{88}, 055025 (2013)
[erratum: Phys. Rev. D \textbf{92}, no.3, 039906 (2015)]
doi:10.1103/PhysRevD.88.055025
[arXiv:1306.4710 [hep-ph]].

%%%


\end{thebibliography}
\end{document}